\documentclass[12pt]{article}
\usepackage[centertags]{amsmath}
\usepackage{graphicx}
\usepackage{upgreek}
\usepackage{amssymb}
\usepackage[makeroom]{cancel}

\begin{document}

\title{Bound states in the Brillouin zone continuum}
\author{ Max D. Porter, Aaron Barr,  Ariel Barr and L.  E. Reichl\\
Center for Complex Quantum Systems and Department of Physics\\
The University of Texas at Austin, Austin, Texas 78712\\}

\date{\today}

\maketitle\

\begin{abstract}

Systems with  space periodic Hamiltonians have unique scattering properties. The discrete translational symmetry associated with periodicity of the Hamiltonian creates scattering channels that govern the scattering process. We  consider a two-dimensional  scattering system in which one dimension is  a periodic lattice and the other is localized in space.  The scattering and decay processes can then be described in terms of  channels indexed by the Bloch momentum. We find the 1D periodic lattice can sustain two types of bound states in the positive energy continuum (BICs): one protected by reflection symmetry, the other protected by discrete translational symmetry. The lattice also sustains long-lived quasibound states. We expect that our results can be generalized to the behavior of states in the continuum of 2D periodic lattices.

\end{abstract}

 \vspace{0.2cm}

\section{Introduction}
\label{sec:intro}

The wave dynamics of nanoscale electronic and photonic condensed matter devices is of interest for a variety of applications.  Periodic lattices are playing  an increasingly important role in electron and photon-based communication because of their capacity for selective wave transmission.  Two-dimensional crystals have exhibited a variety of nanoelectronics applications \cite{Withers,Wu,Ji, Shahrokhi,Song}.  Optical lattices, which can support the propagation of atomic matter waves,  can be formed by superposing multiple pairs of counter-propagating laser beams  \cite{bloch05,lewen07,hemm, greiner,bloch12}.    Light-based communication systems  have been developed using photonic crystals  \cite{Wada,Lyubchanskii,Moniem,Joanno}.    These small nanoscale  quantum and electromagnetic  devices are open systems which can be excited by external waves and subsequently decay.

In recent years,  there has also been increasing interest in locating bound states  that can form in the energy continuum of open systems  (BICs). The first proposal of BICs was due to von Neumann and Wigner in 1929 \cite{vonN,Simon,Stillinger}.   Symmetry-protected quantum BICs,  have been  seen in studies of quantum wires \cite{Schult,Exner,Moiseyev1,Moiseyev2} and electrons in potential surfaces with antisymmetric couplings \cite{Cederbaum}. They were also created using defects attached symmetrically to a 1D waveguide or lattice array \cite{Ladron}, and found to exist due to coupling of materials with different dimensions and different symmetries \cite{Takeichi}. Many other BICs with different confinement mechanisms exist as described in the excellent review paper \cite{Hsu}.
Although a variety of systems have been predicted or observed to have bound states co-existing with continuous propagating waves, much of the focus of the emerging field has been on photonics, due to the relative ease of manipulating photonic structures. 

In subsequent sections, we  consider a potential  with two space dimensions (2D), one an infinite  spatially periodic lattice of potential wells (along the x-axis), the other a Gaussian potential well  (along the z-axis) with effectively finite width. Incident waves can have components that are parallel (along the x-axis) and transverse (along the z-axis) to the 1D periodic lattice.   Because the lattice is spatially periodic, the scattering dynamics is governed by Bloch theory. In subsequent sections, we use a theory originally developed by Wigner and Eisenbud (W-E) \cite{Wigner,Lane,Barrett}, and a method due to Bagwell   \cite{Bagwell} to locate BICs of this 2D system. Although we consider a 1D lattice localized in a 2D plane, an extra lattice dimension would not destroy the  type of BIC we have identified  on the 1D lattice, making the results highly relevant to the study of 2D materials, possibly including graphene and other atomically thin monolayers. 

In Sect.~\ref{sec:potential}, we describe the spatial structure of the scattering system considered here.   In Sect.~\ref{sec:scattering}, we discuss the structure of the scattering problem taking account of the translational symmetry of the lattice.  In Sect.~\ref{sec:W-E}, we build the reaction matrices and then the scattering matrix for the Bloch channels using W-E theory.  In Sect.~\ref{sec:eigen1}, we use reaction matrices to construct an eigenvalue equation that allows us to locate some of the Brillouin zone BICs. In Sect.~\ref{sec:GammaPt}, we show that for a lattice with unit cell reflection symmetry along the lattice dimension the Gamma point has several BICs. In Sect.~\ref{sec:conclude}, we make some concluding remarks.

\section{The Periodic Lattice}
\label{sec:potential}

We consider the dynamics of particles of mass $m$ and energy $E$ in the presence of a periodic lattice which is infinitely long in the $x$-direction and of finite width in the $z$-direction (see Fig. \ref{fig:Lattice}).  The particle dynamics, in the presence of this lattice,  is governed by the Schrodinger equation

\begin{eqnarray}
-\frac{{\hbar}^2}{2m}\left( \frac{d^2}{dx^2}+\frac{d^2}{dz^2} \right){\Upsilon}_E(x,z)+V(x,z){\Upsilon}_E(x,z)=E{\Upsilon}_E(x,z)
\label{Schrod1}
\end{eqnarray}
where ${\Upsilon}_E(x,z)$ is an energy eigenstate with energy $E$. 
The potential energy $V(x,z)$ along the x-axis is given by elliptic theta functions, $\vartheta _3$, and along the z-axis by a single Gaussian potential well, so that
\begin{eqnarray}
V_{\beta}(x,z)=-\frac{U}{2}
\vartheta _3\left( \frac{\pi x}{2a },e^{-\frac{{\epsilon}^2 \pi ^2}{2a^2 }}\right)\frac{a}{\sqrt{2{\pi}}{\sigma}}{\rm exp}\left( \frac{-z^2}{2{\sigma}^2}\right)~~~~~\nonumber\\
-{\beta}\frac{U}{2}
\vartheta _3\left( \frac{\pi (x-a\sqrt{\frac{1}{17}})}{2a },e^{-\frac{{\epsilon}^2 \pi ^2}{2a^2 }}\right)\frac{a}{\sqrt{2{\pi}}{\sigma}}{\rm exp}\left( \frac{-(z+a\sqrt{\frac{1}{13}})^2}{2{\sigma}^2}\right),
\label{PotEn1}
\end{eqnarray}
where $2a$ is the width of the unit cell and ${\sigma}$ and ${\epsilon}$ are standard deviations. 
The elliptic theta functions provide a periodic array of Gaussian potential wells whose shape can be varied by varying the parameter ${\epsilon}$.  In the limit ${\epsilon}{\rightarrow}0$ the periodic array of Gaussians approaches a periodic array of Dirac delta functions (a Dirac comb). For  ${\epsilon}\gtrsim0.5a$ it can be approximated by a convergent sum of cosines \cite{Porter2} and provides a means to model the behavior of quantum particle waves in optical lattices.  More generally, the elliptic theta functions provide a simple and efficient means to model infinite periodic lattices, in any dimension, with a variety of unit cell structures.  In subsequent sections, we use $U=30$,  ${\epsilon}=0.4a$ and ${\sigma}=0.4a$, which models a lattice with potential energy very localized in the unit cells.

Eq.~(\ref{PotEn1}) contains two lines of Gaussian potentials, one with Gaussian potentials centered at $(x=2na, z=0)$ with integer  $(-{\infty}{\leq}n{\leq}\infty)$ and the other with Gaussian potentials centered at irrational positions $(x=a\sqrt{\frac{1}{17}}+2na, z=-a\sqrt{\frac{1}{13}})$. When ${\beta}=0$, only the line of Gaussians centered at $(x=2na, z=0)$, which has unit cell reflection symmetry around both axes,  contributes. However, when ${\beta}{\neq}0$, the reflection symmetry is broken and, as we shall show, there is a qualitative change in the scattering dynamics.

Wigner-Eisenbud theory, which will be described below, requires separation of the system into a  ``reaction region" ($-{\infty}{\leq}x{\leq}{\infty},~-L{\leq}z{\leq}L$), which fully contains the periodic potential, and an asymptotic region ($-{\infty}{\leq}x{\leq}{\infty}$, $L{\leq}|z|$) in which the potential energy $V(x,z)$ is effectively zero. The reaction region and the asymptotic region are then coupled. A plot of five unit cells of the potential energy in the reaction region is given in Fig. 1 for the case ${\beta}=0$  with $L=3a$ and $a=1$.

For an infinitely long periodic lattice, the Bloch momentum is a continuous variable.  However, if we assume the lattice has a finite number $N$ of unit cells, and assume periodic boundary conditions, then we can focus attention on the subset of Bloch momenta $K_{\ell}=\frac{{\ell}{\pi}}{Na}$, $0{\leq}{\ell}{<}N$ (we restrict $\ell$ to states in the Brillouin zone). In the limit $N{\rightarrow}\infty$, we retrieve the continuum of Bloch momenta that determine the dynamical properties of the system.  However, because Bloch momentum is conserved by the dynamics, it is sufficient to consider the finite subset $K_{\ell}=\frac{{\ell}{\pi}}{Na}$, $0{\leq}{\ell}{\leq}N$. In the reaction region, the energy eigenstates decompose into superpositions of Bloch states
{\begin{eqnarray}
{\Upsilon}_{K_{\ell},E}(x,z)={\rm e}^{iK_{\ell}x}{\Phi}_{\ell}(x,z),
\end{eqnarray}
where ${\Phi}_{\ell}(x,z)={\sum}_{\nu=-\infty}^{\infty}{\phi}_{\nu,\ell}(z){\rm e}^{i{\nu}{\pi}x/a}$ has the periodicity of the unit cell.

In the asymptotic region, the Bloch states  take the simple  form
{\begin{eqnarray}
{\Upsilon}^{\alpha}_{{\ell},\nu}(x,z)={\rm e}^{iK_{\ell}x}{\rm e}^{i{\nu}{\pi}x/a}{\xi}^{\alpha}_{k^{\ell}_{\nu}}(z),
\end{eqnarray}
where 
${\xi}^{\alpha}_{k^{\ell}_{\nu}}(z)=A^{\alpha,\ell}_{\nu}{\rm e}^{ik^{\ell}_{\nu}z} +B^{\alpha,\ell}_{\nu}{\rm e}^{-ik^{\ell}_{\nu}z} $ and ${\alpha}=T,B$ distinguishes states for positive (T) and negative (B) z.  
The energy eigenvalues in the asymptotic region  are then given by
{\begin{eqnarray}
E=\frac{{\hbar}^2}{2m}\left(( k^{\ell}_{\nu})^2+\left(K_{\ell}+\frac{{\pi}{\nu}}{a}\right)^2\right)
\end{eqnarray}
where $-{\infty}{\leq}{\nu}{\leq}{\infty}$ denotes the transverse modes for states with Bloch momentum $K_{\ell}$.  Note that the momentum of the particle along the z-axis is then given by 
{\begin{eqnarray}
k^{\ell}_{\nu}=\sqrt{\frac{2m}{{\hbar}^2}E- \left(K_{\ell}+\frac{{\pi}{\nu}}{a}\right)^2}.
\end{eqnarray}
Once the Bloch momentum (the value of $\ell$) is fixed then the scattering channel associated with that Bloch momentum has an infinite number of  scattering modes ${\nu}$ associated with it since $-\infty{\leq}\nu{\leq}\infty$. For finite energy, most of the  scattering modes will be evanescent, but some may be propagating (we define these terms in Sect.~\ref{sec:scattering}). 

We will primarily consider the case $\beta=0$ so that the potential energy unit cell along the x-direction has reflection symmetry about $x=0$. The reflection symmetry of the periodic potential prevents one class of states in the positive energy continuum from decaying, while another class of states is prevented due to conservation of Bloch momentum.   We will  show that when ${\beta}{\neq}0$ new decay paths can open for some of these states. The Schrodinger equation in Eq.~(\ref{Schrod1}) is written in terms of dimensioned  variables, however in subsequent sections we will define various quantities in terms of atomic units. This means $\hbar=1$, and we choose $m=m_e=1,\, a=a_{Bohr}=1,$ and $L=3a$, though often these will be written explicitly for clarity. These dimensionless (atomic) units are used in Figs. 1-11.

\section{Propagating and Evanescent Modes}
\label{sec:scattering}

Particle waves that are incident on the lattice, from either side (T or B), will either be reflected from the lattice or will be transmitted across the lattice along the $z$-axis. We let 
\begin{eqnarray}
{\Upsilon}^{B}_{{\ell},\nu}(x,z)=\left( \frac{1}{\sqrt{k^{\ell}_{\nu}}}A^{{\ell},B}_{\nu}{\rm e}^{ik^{\ell}_{\nu}z}+\frac{1}{\sqrt{k^{\ell}_{\nu}}}B^{\ell,B}_{\nu}{\rm e}^{-ik^{\ell}_{\nu}z} \right){\rm e}^{iK_{\ell}x}{\rm e}^{i{\nu}{\pi}x/a}
\label{incidentB}
\end{eqnarray}
denote the incident $(A^{{\ell},B}_{\nu})$ and reflected $(B^{\ell,B}_{\nu})$ waves in asymptotic region for negative $z$ and let 
\begin{eqnarray}
{\Upsilon}^{T}_{{\ell},\nu}(x,z)=\left( \frac{1}{\sqrt{k^{\ell}_{\nu}}}C^{{\ell},T}_{\nu}{\rm e}^{-ik^{\ell}_{\nu}z}+\frac{1}{\sqrt{k^{\ell}_{\nu}}}D_{\nu}^{{\ell},T}{\rm e}^{ik^{\ell}_{\nu}z} \right){\rm e}^{iK_{\ell}x}{\rm e}^{i{\nu}{\pi}x/a}
\label{incidentT}
\end{eqnarray}
denote the incident $(C^{{\ell},T}_{\nu})$ and reflected $(D_{\nu}^{{\ell},T})$ waves in asymptotic region for positive $z$. 

The details concerning the transmission and reflection of incident waves are contained in the reaction matrices ${\mathcal R}^{{\alpha},{\alpha}'}_{{\nu}_1,{\nu}_2}$ \cite{Lane},  which are defined
\begin{eqnarray}
{\Upsilon}^{\alpha}_{{\ell},\nu_1}(x,z_{\alpha})=-{\sum}_{\nu_2}{\mathcal R}^{\alpha,B}_{{\nu}_1,{\nu}_2}\frac{d{\Upsilon}^{B}_{{\ell},\nu_2}}{dz}{\bigg|}_{z=-L}+{\sum}_{\nu_2}{\mathcal R}^{\alpha,T}_{{\nu}_1,{\nu}_2}\frac{d{\Upsilon}^{T}_{{\ell},\nu_2}}{dz}{\bigg|}_{z=L}
~\label{defreac}
\end{eqnarray}
where ${\alpha}={B,T}$ and ${z_B=-L, z_T=L}$. These equations allow us to relate the incident and reflected waves on both sides (along the z-direction) of the lattice. 

We can now write Eqs.~(\ref{incidentB}), (\ref{incidentT}), and (\ref{defreac})  in matrix form.  To keep the expressions simple, first define
${\mathcal K}^{{\ell},{\alpha}{\alpha '}}_{\nu_1,\nu_2}=~\sqrt{k^{\ell}_{\nu_1}}R^{{\ell},{\alpha}{\alpha '}}_{\nu_1,\nu_2}\sqrt{k^{\ell}_{\nu_2}}$.  Then, combining these equations we can write
\begin{eqnarray}
{\bigg[}\left(
\begin{array}{cc}
{\bar {\bar 1}}_{\ell}  & {\bar {\bar 0}}_{\ell}\\
{\bar {\bar 0}}_{\ell} & {\bar {\bar 1}}_{\ell} \\
\end{array}
\right)+i \left(
\begin{array}{cc}
{\bar  {\bar{\mathcal K}}_{{\ell},BB}} &{\bar  {\bar{\mathcal K}}_{{\ell},BT}} \\
{\bar {\bar{\mathcal K}}_{{\ell},TB} } &{\bar  {\bar{\mathcal K}}_{{\ell},TT} } \\
\end{array}
\right)  {\bigg]}{\cdot} \left(
\begin{array}{cc}
{\bar {\bar U}}_{\ell}&{\bar {\bar 0}}_{\ell} \\
{\bar {\bar 0}}_{\ell}&{\bar {\bar U}}_{\ell} \\
\end{array}
\right) {\cdot}\left(
\begin{array}{c}
{\bar B}_{\ell} \\
{\bar D}_{\ell}\\
\end{array}
\right)~~~~~~~~~~~~~\nonumber\\
= -{\bigg[}\left(
\begin{array}{cc}
{\bar {\bar 1}}_{\ell}&{\bar {\bar 0}}_{\ell} \\
{\bar {\bar 0}}_{\ell}&{\bar {\bar 1}}_{\ell} \\
\end{array}
\right) -i \left(
\begin{array}{cc}
{\bar {\bar {\mathcal K}}}_{{\ell},BB} & {\bar {\bar { \mathcal K}}}_{{\ell},BT} \\
{\bar {\bar {\mathcal K}}}_{{\ell},TB}  &{\bar  {\bar {\mathcal K}}}_{{\ell},TT}  \\
\end{array}
\right)  {\bigg]}{\cdot} \left(
\begin{array}{cc}
{\bar {\bar U}^{\dagger}}_{\ell}&{\bar {\bar 0}}_{\ell} \\
{\bar {\bar 0}}_{\ell}&{\bar {\bar U}^{\dagger}}_{\ell} \\
\end{array}
\right) {\cdot} \left(
\begin{array}{c}
{\bar A}_{\ell} \\
{\bar C}_{\ell} \\
\end{array}
\right)~
\label{Smat1}
\end{eqnarray}
where ${\bar A}_{\ell}$, ${\bar B}_{\ell}$, ${\bar C}_{\ell}$, and ${\bar D}_{\ell}$ are infinite dimensional column matrices,  ${\bar {\bar 0}}_{\ell}$ is an infinite dimensional  square matrix with all matrix elements equal to zero, and the matrices ${\bar  {\bar{\mathcal K}}_{{\ell},{\alpha},{\alpha '}}} $ are infinite dimensional square matrices whose matrix elements are ${\mathcal K}^{{\ell},{\alpha}{\alpha '}}_{\nu_1,\nu_2}$.  ${\bar {\bar U}}_{\ell}$ is an infinite dimensional diagonal matrix with diagonal matrix elements ${\rm e}^{ik^{\ell}_{{\nu}}L}$,  where $-\infty{\leq}{\nu}{\leq}\infty$.    Since $\ell$ is fixed but $-\infty{\leq}{\nu}{\leq}\infty$,  each $\ell$ has an infinite-dimensional matrix equation (although we will discuss how to approximate these as finite-dimensional for finite energy). 
If we multiply by the inverse of the matrices on the left, 
we finally obtain
{\begin{eqnarray}
\left(
\begin{array}{c}
{\bar B}_{\ell}\\
{\bar D}_{\ell}\\
\end{array}
\right)={\bar {\bar S}}^{(\ell)}{\cdot}\left(
\begin{array}{c}
{\bar A}_{\ell} \\
{\bar D}_{\ell}\\
\end{array}
\right)
\end{eqnarray}
where 
{\begin{eqnarray}
{\bar {\bar S}}^{(\ell)}=-{\bar {\bar U}}_{\ell}^{\dagger}{\cdot}[{\bar {\bar 1}}_{\ell}+i{\bar {\bar K}}_{\ell}]^{-1}{\cdot}[{\bar {\bar 1}}_{\ell}-i{\bar {\bar K}}_{\ell}]{\cdot}{\bar {\bar U}}_{\ell}^{\dagger}
\end{eqnarray}
is the scattering matrix and contains information about transmission and reflection of particles approaching the lattice from either side.

We now must distinguish between propagating and evanescent scattering modes \cite{Bagwell}. 
For  energy $E$ and Bloch momentum $K_\ell=\frac{\ell \pi}{N}$, propagating modes  are modes $\nu$ with real values of the wave vector 
$k^{\ell}_{\nu}=\sqrt{2E- \left(\frac{{\ell}{\pi}}{N}+ {\nu}{\pi}\right)^2}$, so that $E{\geq}\frac{1}{2} \left(\frac{{\ell}{\pi}}{N}+ {\nu}{\pi}\right)^2$.  
Evanescent modes are modes $\nu$ with energy $E{<}\frac{1}{2} \left(\frac{{\ell}{\pi}}{N}+ {\nu}{\pi}\right)^2$. In that case, the wave vector is defined $k^{\ell}_{\nu}=+iq^{\ell}_{\nu}$, where \\ $q^{\ell}_{\nu}=\sqrt{ \left(\frac{{\ell}{\pi}}{N}+ {\nu}{\pi}\right)^2-2E}$. With this choice of wave vector for the evanescent modes, the outgoing modes decay exponentially and incoming modes grow exponentially as we move away from the reaction region. Thus,  we set the amplitudes of the incoming modes to zero so that $A^{{\ell},B}_{\nu_1}=0$ and $C^{{\ell},T}_{\nu_1}=0$.  This ensures that evanescent modes decay away from the reaction region.  When evanescent modes are included, ${\bar {\bar S}}^{(\ell)}$ is not a unitary matrix.  However, the elements of ${\bar {\bar S}}^{(\ell)}$ that connect the incident propagating modes to the outgoing propagating modes do  form a unitary matrix and can be identified as the scattering matrix for the propagating modes.

As an example of how these equations are used, let us consider the case $\ell=0$. This corresponds to scattering in the $\Gamma$  point Bloch channel of the Brillouin zone. 
We consider the energy interval $0{\leq}E{\leq}\frac{1}{2}\pi^2$, which is the lowest energy scattering mode for this channel.  Then there is only one propagating mode  ($\nu=0$) and an infinite number of evanescent modes ($\nu{\neq}0$). We will keep only the slowest decaying evanescent modes $\nu={\pm}1$. Then $k^{0}_{0}=\sqrt{2E}$
and $k^{0}_{\pm1}=iq^{0}_{\pm1}$ where $q^{0}_{\pm1}=\sqrt{ \pi^2-2E}$. 
The matrix equation now takes the form
{\begin{eqnarray}
\left(
\begin{array}{c}
B^{{0},B}_{-1}\\
B^{{0},B}_{0}\\
B^{{0},B}_{+1}\\
D^{{0},T}_{-1}\\
D^{{0},T}_{0}\\
D^{{0},T}_{+1}\\
\end{array}
\right)={\bar {\bar S}}^{(0)}{\cdot}\left(
\begin{array}{c}
0\\
A^{{0},B}_{0}\\
0\\
0\\
C^{{0},T}_{0}\\
0\\
\end{array}
\right)
\label{Sevan}
\end{eqnarray}
In this case, the sub-matrix 
{\begin{eqnarray}
\left(
\begin{array}{cc}
{\bar {\bar S}}^{(0)}_{2,2} &{\bar {\bar S}}^{(0)}_{2,5}\\
{\bar {\bar S}}^{(0)}_{5,2} &{\bar {\bar S}}^{(0)}_{5,5}\\
\end{array}
\right)
\end{eqnarray}
is unitary and is the scattering matrix for the propagating modes.

\section{Reaction Matrix for the Bloch Channels}
\label{sec:W-E}

We can derive the reaction matrix for the various Bloch channels using Wigner-Eisenbud (W-E)  theory.  W-E theory (also called reaction matrix theory) decomposes configuration space into a reaction region  that fully contains the  energy and an asymptotic  region in which the potential energy is effectively zero.  The reaction region   can be described in terms of a complete set of basis states that satisfy certain boundary conditions on the interface between the reaction and asymptotic regions \cite{Lane}.  The reaction region is  coupled to the asymptotic region with a singular coupling \cite{Bloch, Feshbach}. It is useful to note that  W-E scattering  theory more recently  has also been used to analyze electron scattering in quasi-one-dimensional waveguides \cite{Akguc, Lee}, for analyzing the dissociation of the HOCl molecule \cite{Barr2}, and for scattering  of waves from systems with cylindrical symmetry \cite{Barr3,reichl-1}. The procedure is well described in \cite{reichl-2,Akguc-2}, so we only outline the steps here.

We can separate the eigenvalue equations into contributions from the reaction region and asymptotic regions using projection operators.  Let 
\begin{eqnarray}
{\hat Q}={\int_{-L}^{L}}dz{\int_{-Na}^{Na}}dx~|x,z{\rangle}{\langle}x,z|,~~{\rm and}~~{\hat P}_{\alpha}={\int_{\alpha}}dz{\int_{-Na}^{Na}}dx~|x,z{\rangle}{\langle}x,z|~,~~
\end{eqnarray}
where ${\alpha}={\{}B,T{\}}$,  ${\int}_B{\equiv}{\int}_{-\infty}^{-L}$, ${\int}_T{\equiv}{\int}_{L}^{\infty}$, and ${\langle}x',z'|x,z{\rangle}={\delta}(x'-x){\delta}(z'-z)$.
It is straightforward to show that ${\hat P}_B{\hat P}_B={\hat P}_B$, ${\hat P}_T{\hat P}_T={\hat P}_T$, ${\hat Q}{\hat Q}={\hat Q}$, and ${\hat P}_B{\hat Q}={\hat P}_T{\hat Q}=0$ and that ${\hat P}_B+{\hat P}_T+{\hat Q}={\hat 1}$.

Projections of the eigenvalue equation, ${\hat H}|E{\rangle}=E|E{\rangle}$ (Eq.~(\ref{Schrod1})) take the form
\begin{eqnarray}
{\hat Q}{\hat H}{\hat Q}|E{\rangle}+{\sum}_{\alpha}{\hat Q}{\hat H}{\hat P}_{\alpha}|E{\rangle}=E{\hat Q}|E{\rangle}~\label{EigEq}\\
{\hat P}_{\alpha}{\hat H}{\hat Q}|E{\rangle}+{\sum}_{\alpha '}{\hat P}_{\alpha}{\hat H}{\hat P}_{\alpha '}|E{\rangle}=E{\hat P}_{\alpha}|E{\rangle}~
\end{eqnarray}
The coupling Hamiltonians ${\hat Q}{\hat H}{\hat P}_B$ and  ${\hat Q}{\hat H}{\hat P}_T$  are defined
\begin{eqnarray}
{\hat Q}{\hat H}{\hat P}_B=-\frac{2{\hbar}^2}{m}{\hat Q}{\delta}(z+L){\partial}^{\rightarrow}_z{\hat P}_B~~{\rm and}~~{\hat Q}{\hat H}{\hat P}_T=\frac{2{\hbar}^2}{m}{\hat Q}{\delta}(z-L){\partial}^{\rightarrow}_z{\hat P}_T
\end{eqnarray}
In Appendix A, we show how to compute matrix elements of these coupling Hamiltonians (see also \cite{reichl-2}).

\subsection{Reaction Region Eigenstates}
\label{sec:eigen}

The eigenvalue problem in the reaction region can be written
\begin{eqnarray}
{\hat H}_{QQ}{\hat Q}|{\phi}_{j} {\rangle}={\lambda}_{j}{\hat Q}|{\phi}_{j}{\rangle}
\end{eqnarray}
 where ${\hat H}_{QQ}={\hat Q}{\hat H}{\hat Q}$ and  ${\hat H}=\frac{1}{2m}({\hat p}_x^2+{\hat p}_y^2)+V(x,z)$.  
 The eigenstates ${\hat Q}|{\phi}_{j}{\rangle}
$ are orthonormal and complete so ${\langle}{\phi}_{j}|{\hat Q}|{\phi}_{j'}{\rangle}={\delta}_{{j},{j '}}$  and ${\sum_{{j}=0}^{\infty}}{\hat Q}|{\phi}_{{j}}{\rangle}{\langle}{\phi}_{{j}}|{\hat Q}={\hat Q}$.

The eigenstates in the $N$-cell reaction region can be written
\begin{equation}
{\langle}x,z|{\hat Q}|{\phi}_{j}{\rangle}={\sum_{m={-\infty}}^{\infty}}{\sum_{n=0}^{\infty}}C^{j}_{m,n}{\langle}x,z|m,n{\rangle}
\end{equation}
where
\begin{eqnarray}
{\langle}x,z|m,n{\rangle}=
\frac{1}{\sqrt{2Na}}{\rm e}^{ik_mx}{\xi}_{n}(z)
\end{eqnarray}
for $k_m=\frac{{\pi}m}{Na}$, 
$-{\infty}{\leq}m{\leq}{\infty}$, and
\begin{eqnarray}
{\xi}_{n}(z)=\left\{ \begin{array}{ll}
\sqrt{\frac{1}{2L}},~~&{\rm for~ n}=0, \\
\sqrt{\frac{1}{L}}{\cos}\left[\frac{n{\pi}(z+L)}{2L}\right]~~&{\rm for}~ n=1,...,{\infty},
\end{array} \right.
\end{eqnarray}
is a complete orthonormal basis.
These states are normalized over the interval $-Na{\leq}x{\leq}Na$ and $-L{\leq}z{\leq}L$, and $C^{j}_{m,n}={\langle}m,n|{\hat Q}|{\phi}_{j}{\rangle}$. 

The eigenvalue equation in the reaction region then takes the form 
\begin{eqnarray}
{\langle}x,z|{\hat Q}{\hat H}{\hat Q}|x,z{\rangle}{\langle}x,z|{\hat Q}|{\phi}_{j} {\rangle}=-\frac{{\hbar}^2}{2m}\left( \frac{d^2}{dx^2}+\frac{d^2}{dz^2} \right) {\langle}x,z|{\hat Q}|{\phi}_{j} {\rangle}~~~~~\nonumber\\
+V(x,z){\langle}x,z|{\hat Q}|{\phi}_{j} {\rangle}={\lambda}_j  {\langle}x,z|{\hat Q}|{\phi}_{j} {\rangle}.
\label{eigeq1}
\end{eqnarray}
The eigenstates ${\langle}x,z|{\hat Q}|{\phi}_{j}{\rangle}$ will naturally decompose into groupings of states, each with a different Bloch momentum $K_{\ell}=\frac{{\ell}{\pi}}{Na}$, $0{\leq}{\ell}{<}N$ (we restrict $\ell$ to states in the Brillouin zone). The collection of eigenstates, ${\langle}x,z|{\hat Q}|{\phi}^{\ell}_{j}{\rangle}$,  with  Bloch momentum $K_{\ell}=\frac{{\ell}{\pi}}{Na}$, can be written in the form
\begin{equation}
{\langle}x,z|{\hat Q}|{\phi}^{\ell}_{j}{\rangle}={\rm e}^{iK_{\ell}x}\frac{1}{\sqrt{2Na}}{\sum_{m={-\infty}}^{\infty}}{\sum_{n=0}^{\infty}}C^{j,{\ell}}_{m,n}{\rm e}^{im{\pi}x/a}{\xi}_{n}(z).
\end{equation}
so that the collection of $N$-cell eigenstates $\{{\langle}x,z|{\hat Q}|{\phi}_{j}{\rangle}\}$ decomposes into $N$ disjoint sets of states, each with a different Bloch momentum $K_{\ell}$.

\subsection{Reaction Matrices}
\label{sec:ReactionMat}

We now  construct the reaction matrices.  The first step is to expand the exact energy eigenstates of the open system, ${\langle}x,z|E{\rangle}$, in terms of reaction region Bloch eigenstates  ${\langle}x,z|{\hat Q}|{\phi}^{\ell}_{j} {\rangle}$ and asymptotic region eigenstates ${\langle}x,z|{\hat P}_{\alpha}|{\psi}^{\alpha}_{k^{\ell}_{\nu},\nu}{\rangle}=\Upsilon_{\ell,\nu}^{\alpha}(x,z)$. Then the exact energy eigenstates take the form
\begin{eqnarray}
|E{\rangle}={\sum_{\ell=0}^{N-1}}~{\sum_{j =1}^{\infty}}~{\gamma}^{\ell}_{j}(E){\hat Q}|{\phi}^{\ell}_{j}{\rangle}+{\sum_{\alpha}}~{\sum_{\ell=0}^{N-1}}~{\sum_{\nu=-{\infty}}^{\infty}}~{\Gamma}^{\alpha}_{{k^{\ell}_{\nu},\nu}}(E){\hat P}_{\alpha}|{\psi}^{\alpha}_{k^{\ell}_{\nu},\nu}{\rangle}
\end{eqnarray}
where ${\gamma}^{\ell}_{{j}}(E)={\langle}{\phi}^{\ell}_{{j}}|{\hat Q}|E{\rangle}$ and 
${\Gamma}^{\alpha}_{{ k^{\ell}_{\nu},\nu}}(E)={\langle}{\psi}^{\alpha}_{k^{\ell}_{\nu},\nu}|{\hat P}_{\alpha}|E{\rangle}$ are complex coefficients that determine the weight of the contribution of each of the basis states  ${\hat Q}|{\phi}^{\ell}_{j} {\rangle}$ and ${\hat P}_{\alpha}|{\psi}_{k^{\ell}_{\nu},\nu}^{\alpha}{\rangle}$ to the energy eigenstates $|E{\rangle}$ of the system.

To obtain the reaction matrix, we consider again Eq.~(\ref{EigEq}), 
and note that the spectral decomposition of ${\hat Q}{\hat H}{\hat Q}$ is 
\begin{eqnarray}
{\hat Q}{\hat H}{\hat Q}={\sum_{\ell=0}^{N-1}}~{\sum_{j=1}^{\infty}}{\lambda}^{\ell}_{j} {\hat Q}|{\phi}^{\ell}_{j}{\rangle}{\langle}{\phi}^{\ell}_{j}|{ \hat Q}.
\end{eqnarray}
We can therefore write Eq. (\ref{EigEq}) in the form
\begin{eqnarray}
{\sum_{\ell=0}^{N-1}}~{\sum_{j}}{\lambda}^{\ell}_{j} {\hat Q}|{\phi}^{\ell}_{j}{\rangle}{\langle}{\phi}^{\ell}_{j}|{ \hat Q}|E{\rangle}+{\sum_{\alpha}}{\hat Q}{\hat H}{\hat P}_{\alpha}|E{\rangle}=E{\hat Q}|E{\rangle}~~~~
\label{Qeig1}
\end{eqnarray}
Now multiply by   ${\langle}{\phi}^{{\ell}'}_{j'} |$ to get
\begin{eqnarray}
(E-{\lambda}^{{\ell}}_{j}){\gamma}^{{\ell}}_{j}(E)-{\sum_{\alpha}} {\sum_{\ell '=0}^{N-1}}{\sum_{\nu=-{\infty}}^{\infty}}{\langle}{\phi}^{{\ell}}_{j} |{\hat Q}{\hat H}{\hat P}_{\alpha}|{\psi}_{k^{{\ell}'},\nu}^{\alpha} {\rangle}{\Gamma}_{k^{{\ell}'}_{\nu},\nu}^{\alpha}(E)=0~~~
\label{Qeig1}
\end{eqnarray}
Using the definition of the coupling Hamiltonian ${\hat Q}H{\hat P}_{\alpha}$ given in Appendix A, this reduces to
\begin{eqnarray}
{\gamma}^{\ell}_{j}(E)=-\frac{{\hbar}^2}{2m}  {\sum_{\nu=-{\infty}}^{\infty}}\frac{{\Phi}^{\ell *}_{j,\nu}(-L)}{(E-{\lambda}^{\ell}_{j})} \frac{d{\xi}^B_{k^{\ell}_{\nu}}}{dz}{\bigg|}_{z=-L} {\Gamma}_{k^{\ell}_{\nu}}^{B}(E)~~~~~~~~~~~~~\nonumber\\
+\frac{{\hbar}^2}{2m} {\sum_{\nu=-{\infty}}^{\infty}} \frac{{\Phi}^{\ell *}_{{j},\nu}(L)}{(E-{\lambda}^{\ell}_{j })} \frac{d{\xi}^T_{k^{\ell}_{\nu}}}{dz}{\bigg|}_{z=L} {\Gamma}_{k^{\ell}_{\nu}}^{T}(E),
\label{reaction4}
\end{eqnarray}
where the quantities ${\Phi}^{\ell *}_{j,\nu}(z_{\alpha})$ are defined in Appendix A. It is useful to note that the coupling Hamiltonian ${\hat Q}H{\hat P}_{\alpha}$ is singular for coupling between the reaction region and the asymptotic regions along the $z$-direction. It provides a means to couple the zero-slope eigenstates of the reaction region to the scattering states in such a way that the resulting exact energy eigenstates have a continuous slope.  Along the $x$-direction it provides the mechanism to form Bloch channels due to the orthogonality of Bloch states with different Bloch momentum.

We next impose the condition that the scattering eigenstate ${\langle}x,z|E{\rangle}$ be continuous at $z=z_{\alpha}$ so ${\langle}x,z_{\alpha}|{\hat Q}|E{\rangle}={\langle}x,z_{\alpha}|{\hat P}_{\alpha}|E{\rangle}$.  If we multiply by $\frac{1}{\sqrt{2Na}}{\rm e}^{-iK_{\ell}x}{\rm e}^{-i{\nu}{\pi}x/a}$ and integrate over $x$, we obtain
\begin{eqnarray}
{\sum_{j=1}^{\infty}}{\Phi}^{\ell}_{\nu,j}(z_{\alpha}) {\gamma}^{\ell}_{{j}}(E)={\Gamma}_{k^{\ell}_{\nu}}^{\alpha}(E){\xi}_{k^{\ell}_{\nu}}^{\alpha}(z_{\alpha}).
\label{cont0}
\end{eqnarray}
The reaction matrix is obtained from the condition
\begin{eqnarray}
{\sum_{\alpha '} } {\sum_{{\nu}_2=-\infty}^{\infty}} 
 R^{\ell,\alpha \alpha  '}_{{\nu}_1,{\nu}_2}(E)~ \frac{d{\xi}^{\alpha '}_{k^{\ell}_{\nu_2}}}{dz}{\bigg|}_{z_{\alpha '}}{\Gamma}_{k^{\ell}_{\nu_2}}^{\alpha '}(E)={\Gamma}_{k^{\ell}_{\nu_1}}^{\alpha}(E){\xi}_{k^{\ell}_{\nu_1}}^{\alpha}(z_{\alpha })
\label{reaction6}
\end{eqnarray}

If we combine  Eqs. (\ref{reaction4}) and (\ref{cont0}) and note that $z_L=-L$ and $z_R=+L$,  we obtain 
 the following expression for the reaction matrices
\begin{eqnarray}
R^{{\ell},{\alpha}{\alpha}'}_{\nu_1,\nu_2}(E)=\frac{{\hbar}^2}{2m} {\sum_{j=1}^{\infty}}  \frac{{\Phi}^{\ell}_{\nu_1,j}(z_{\alpha}){\Phi}^{{\ell}*}_{j,\nu_2}(z_{\alpha '})}{(E-{\lambda}^{\ell}_{j })}
\label{reaction7}
\end{eqnarray}
where ${\alpha}=B,T$.

For scattering potentials with a clearly defined boundary  and  for which the scattering problem can be solved analytically, the W-E theory gives exact agreement with the exact results   \cite{Barr3, reichl-2,Akguc-2} (this is shown in Appendix B for scattering from a square-well potential).  
For the case of a scattering potential without a well defined boundary, such as we are considering here, we have found that the reaction region contains two types of states: {\it localized} states that do not depend on the size $L$  of the reaction region, and {\it extended} states whose number and shape depend on $L$.  In Fig. \ref{fig:Ldepend}, we show the behavior of the  reaction region  eigenstates, in the energy interval, $0{\leq}E{\leq}\frac{{\pi}^2}{2}$ (the first scattering mode of the $\Gamma$  point channel), as a function of the location  of the boundary $L$.  We find that there are only two {\it localized} states that are independent of the position of the boundary $L$ and  are intrinsic to the scattering problem.  These are indicated by the two solid blue lines in Fig. \ref{fig:Ldepend} and are shown in Figs. \ref{fig:Ldependfigs}.a and \ref{fig:Ldependfigs}.b for $L=3$, and in Figs. \ref{fig:L5dependfigs}.a and \ref{fig:L5dependfigs}.b.  for $L=5$. These states are well localized in the neighborhood of the lattice potential energy. The remaining states in Figs. \ref{fig:Ldependfigs} and  \ref{fig:L5dependfigs}  clearly see the size of the reaction region and show branching behavior with increasing $L$ where  new states with the same structure, but an additional wavelength appear in the reaction region.  In Figs. \ref{fig:Ldependfigs}.c and \ref{fig:Ldependfigs}.d we show examples of  two of the {\it extended}  L-dependent  states for $L=3$ and in Figs. \ref{fig:L5dependfigs}.c and \ref{fig:L5dependfigs}.d we show two {\it extended} states for $L=5$.  The localized states give rise to very long-lived quasibound states or to bound states in the energy continuum.

\section{Eigenvalue Equation}
\label{sec:eigen1}

Equation~(\ref{reaction6}) is the eigenvalue equation for this system. As shown by Bagwell \cite{Bagwell}, we can use it to search for bound states of the system in regions that only contain evanescent modes (${\nu}=0$). We can write them in a matrix form as

\begin{eqnarray}
\left(\begin{array}{cc}
{\xi}^B_{k^{\ell}_{0}}
-R^{{\ell},BB}_{0,0}(E) \frac{d{\xi}^B_{k^{\ell}_{0}}}{dz}{\bigg|}_{z=-L},& R^{{\ell},BT}_{0,0}(E) \frac{d{\xi}^T_{k^{\ell}_{0}}}{dz}{\bigg|}_{z=L}\\
-R^{{\ell},TB}_{0,0}(E) \frac{d{\xi}^B_{k^{\ell}_{0}}}{dz}{\bigg|}_{z=-L}, &{\xi}^T_{k^{\ell}_{0}}
+R^{{\ell},TT}_{0,0}(E) \frac{d{\xi}^T_{k^{\ell}_{0}}}{dz}{\bigg|}_{z=L}\\
\end{array}\right){\cdot} \left(\begin{array}{c}{\Gamma}_{k^{\ell}_{0}}^{B}(E) \\
 {\Gamma}_{k^{\ell}_{0}}^{T}(E) \\
 \end{array}\right)=0,
\label{Hbd1}
\end{eqnarray}
where 
\begin{eqnarray}
{\xi}^B_{k^{\ell}_{0}}=\frac{1}{\sqrt{q_0^{\ell}}}B_0^B{\rm e}^{-q_0^{\ell}L},~~~{\xi}^T_{k^{\ell}_{0}}=\frac{1}{\sqrt{q_0^{\ell}}}D_0^T{\rm e}^{-q_0^{\ell}L} \nonumber\\
 \frac{d{\xi}^B_{k^{\ell}_{0}}}{dz}{\bigg|}_{z=-L}=-\sqrt{q_0^{\ell}}B_0^B{\rm e}^{-q_0^{\ell}L},~~~ \frac{d{\xi}^T_{k^{\ell}_{0}}}{dz}{\bigg|}_{z=L}=-\sqrt{q_0^{\ell}}D_0^T{\rm e}^{-q_0^{\ell}L}
\label{reactionmat1}
\end{eqnarray}
We can rewrite Eq. (\ref{Hbd1}) in the form
\begin{eqnarray}
\left(\begin{array}{cc}
1+q_0^{\ell}R^{{\ell},BB}_{0,0}(E),& -q_0^{\ell}R^{{\ell},BT}_{0,0}(E)\\
q_0^{\ell}R^{{\ell},TB}_{0,0}(E), &1
-q_0^{\ell}R^{{\ell},TT}_{0,0}(E)\\
\end{array}\right){\cdot} \left(\begin{array}{c}B_0^B{\Gamma}_{k^{\ell}_{0}}^{B}(E) \\
 D_0^T{\Gamma}_{k^{\ell}_{0}}^{T}(E) \\
 \end{array}\right)\frac{{\rm e}^{-q_0^{\ell}L}}{\sqrt{q_0^{\ell}}}=0.
\label{Hbd2}
\end{eqnarray}
Eq. (\ref{Hbd2}) is an eigenvalue equation for the ``bound states" of the system. The bound state energies are given by the condition that the determinant of the 2x2 matrix 
\begin{eqnarray}
H_{bd}=\left(\begin{array}{cc}
1+q_0^{\ell}R^{{\ell},BB}_{0,0}(E),& -q_0^{\ell}R^{{\ell},BT}_{0,0}(E)\\
q_0^{\ell}R^{{\ell},TB}_{0,0}(E), &1
-q_0^{\ell}R^{{\ell},TT}_{0,0}(E)\\
\end{array}\right)
\label{Hbd3}
\end{eqnarray}
be zero (${\rm Det}[H_{bd}]=0$).
Some of these ``bound states" may have positive energy because they have no propagating path to escape the region of the lattice. 

In Fig. \ref{fig:BlochBSspectrum}, we plot ${\rm Det}[H_{bd}]$ as a function of energy for the scattering channel with Bloch wavevector $K_{\ell}=\frac{\pi}{3}$.  For this case, there are no propagating modes for energies $E<\frac{1}{2}\left(\frac{\pi}{3}\right)^2{\approx}0.5483$. Here we restrict summation over $j$ in the reaction matrix Eq.~(\ref{reaction7}) to states in this energy range (6 states total) plus one evanescent mode.  As we show in Fig. \ref{fig:BlochBSspectrum}, we find negative energy solutions and one positive energy solution, indicating that, for this Bloch channel the lattice has a bound state in the positive energy continuum. We find similar behavior for the Bloch channel $K_{\ell}=\frac{2\pi}{5}$. Plots of these two Bloch BICs are shown in Fig. \ref{fig:BlochBSplots}.  In fact as shown in Fig.  \ref{fig:bandasymp} there is a whole line of positive energy Bloch states that form BICs because, although they have positive energy, their energy lies below that of the first propagating mode and their energy is evanescent. It is interesting that states with similar behavior have been seen in electromagnetic systems, either as evanescent waves that propagate along linear arrays of spheres \cite{Linton} or in the presence of dielectric gratings \cite{Bulgakov}.

\section{BICs at the ${\Gamma}$-point}
\label{sec:GammaPt}

As described above, there is a line of BICs with finite Bloch momentum that exist because at their Bloch momentum there is no propagating mode available to them, and propagating modes in other Bloch channels cannot couple to them.  They exist regardless of whether or not the periodic  lattice unit cell has reflection symmetry.   At the $\Gamma$  point,  there is another type of BIC that can exist for the case ${\beta}=0$. The lattice with ${\beta}=0$ has reflection symmetry about $x=0$ and this symmetry prevents some states from coupling to the continuum and decaying. 
We show the effect of this for the case $0{\leq}E{\leq	}\frac{{\pi}^2}{2}$ where one propagating mode exists, and the case $\frac{{\pi}^2}{2}{\leq}E{\leq}\frac{4{\pi}^2}{2}$, where three propagating modes exist.

\subsection{$\Gamma$  point first energy channel -  $0{\leq}E{\leq}\frac{{\pi}^2}{2}$ }

Symmetry-protected BICs appear to exist only at the $\Gamma$ point. They occur in the lowest-energy scattering mode of the $\Gamma$ point channel, where the  propagating mode has no transverse component, making it transversely even under reflection. As a result, any transversely odd lattice eigenstate in this energy range cannot couple to it and decay, thereby becoming a bound state. For the energy region  $0{\leq}E{\leq}\frac{{\pi}^2}{2}$ at the $\Gamma$  point, there are only two localized states and they are both antisymmetric functions of $x$ (see Figs. \ref{fig:Ldependfigs}.a and   \ref{fig:Ldependfigs}.b). For ${\beta}=0$, they therefore cannot couple to the continuum and cannot decay. (This is similar to the photonic crystal slab described in \cite{Hsu}, though that system concerns electromagnetic waves not particle waves.) 

Breaking the lattice reflection symmetry will downgrade the two BICs in this channel of the $\Gamma$  point to  long-lived quasibound states.
Indeed, for $\beta=0$, the transmission and reflection amplitudes have no poles in the complex energy plane due to the two localized reaction region states in Fig.  \ref{fig:Ldependfigs}.  For ${\beta}{\neq}0$, two poles emerge in the complex energy plane indicating that the states that are BICs at $\beta=0$ become quasibound states for $\beta{\neq}0$. In Fig. \ref{fig:BiCPoles}, we show these two poles for  $\beta=0.01$.  If we take the limit $\beta\rightarrow 0$, we observe the lifetime of the quasibound states going to infinity, as shown in Fig.~ \ref{fig:BiCLifetime}. More precisely, the first BIC (near $E{\approx}0.6$) has $-{\rm Im}[E]=0.38{\beta}^2$ while the second BIC (near $E{\approx}4.7$) has $-{\rm Im}[E]=0.03{\beta}^2$. Lifetime of a quasibound state is given by $\Gamma=\frac{\hbar}{-\rm Im[E]}$, so both states have lifetimes that scale as $1/\beta^2$, where $\beta$ is the size of the asymmetric perturbation. To get lifetimes in SI units, the conversion is (1 unit of time in atomic units = $2.42*10^{-17}$ sec). In an experiment, this lattice could experience small noisy perturbations and still have impressively long-lived states in the continuum.

\subsection{$\Gamma$  point second energy channel -  $\frac{{\pi}^2}{2}{\leq}E{\leq}\frac{4{\pi}^2}{2}$ }

As can be seen in Fig. 2,  in the energy interval $\frac{{\pi}^2}{2}{\leq}E{\leq}\frac{4{\pi}^2}{2}$ at the $\Gamma$  point, there are three propagating modes, corresponding to modes with wavevectors $k_0^0=\sqrt{2E}$, $k_{-1}^0=k_{+1}^0=\sqrt{2E-{\pi}^2}$. In this energy interval there are four localized reaction region eigenstates. These states are shown in Fig. \ref{fig:UpGamStates}. As one might expect from their structure, they are long-lived quasibound states in the continuum (similar states have been called QBICs \cite{Garmon}), and they come in odd-even pairs.

If we include the evanescent modes $k_{-2}^0=k_{2}^0=i\sqrt{4{\pi}^2-2E}$, then the ``S-matrix`` in Eq.~(\ref{Sevan}) is a $10{\times}10$ matrix, which contains in it a $6{\times}6$ unitary scattering matrix that has the form
\begin{eqnarray}
{\bar {\bar S}}^{(0)}=\left(\begin{array}{cc}
{\bar {\bar R}}^{BB}&{\bar {\bar T}}^{BT}\\
{\bar {\bar T}}^{TB}&{\bar {\bar R}}^{TT}\\
\end{array}\right).
\label{subS22}
\end{eqnarray}
where, for example, ${\bar {\bar R}}^{BB}$ is the $3{\times}3$ matrix of reflection amplitudes between the three modes ${\nu}=0,{\pm}1$ for particle waves entering the lattice from below. It can be written
\begin{eqnarray}
{\bar {\bar R}}^{BB}=\left(\begin{array}{ccc}
R_{mm}^{BB}& R_{m0}^{BB}&R_{mP}^{BB}\\
R_{0m}^{BB}& R_{00}^{BB}&R_{0p}^{BB}\\
R_{pm}^{BB}& R_{p0}^{BB}&R_{pp}^{BB}\\
\end{array}\right)
\label{Hbd3}
\end{eqnarray}
where $m$ and $p$ refer to the channels  ${\nu}=-1$ and ${\nu}=+1$, respectively. Here $R_{mm}^{BB}$ is the reflection amplitude for a wave entering the lattice from below in channel ${\nu}=-1$ and then reflecting back into channel ${\nu}=-1$.  

The fact that the localized states \ref{fig:UpGamStates}.a and \ref{fig:UpGamStates}.c are antisymmetric means that they cannot participate in the scattering process for ${\beta}=0$ in mode ${\nu}=0$.  We can see the effect of this on the transmission and reflection coefficients for these modes.  In Fig. \ref{fig:UpGamReflec} we show the reflection coefficients $R_{mm}$, $R_{m0}$, and $R_{00}$, for the energy interval $\frac{{\pi}^2}{2}{\leq}E{\leq}\frac{4{\pi}^2}{2}$, for both ${\beta}=0$ and ${\beta}=0.04$. (Note that we have used only the localized states to construct the reaction matrices used for these plots.) The lack of participation of the lower energy antisymmetric state is particularly striking for the reflection coefficients  $R_{m0}$ and $R_{00}$.  

The evidence that these four states are quasibound comes from the structure of the poles of the S-matrix in the complex energy plane.  
In Fig. \ref{fig:UpGamPoles} we show the poles of $R_{mm}$ and $R_{00}$ for ${\beta}=0$ and ${\beta}=0.04$.  For $R_{mm}$, each localized state is associated with a pole. For $R_{00}$, there are two poles for ${\beta}=0$ (the low energy pole is very tiny) and  much enhanced pole structure for ${\beta}=0.04$. The distance of each pole from the real energy axis is inversely proportional to the state's lifetime. 

If we look at higher energies in the reaction region at the $\Gamma$  point, we find  four more localized states, again in odd-even pairs. Their energies are $E=34.1872, 34.1877$ (odd, even) and $E=41.1521, 41.1556$ (odd, even). We  expect that additional  quasibound state pairs, similar to these,  persist to even higher energies and may be relatively long-lived.

\section{Conclusions}
\label{sec:conclude}

We have found that scattering from lattices with discrete space translation symmetry is governed by scattering channels what conserve Bloch momentum.  The existence of these scattering channels can prevent some positive energy states from decaying, thereby forming bound states in the positive energy continuum. 

 For the case, where the spatially periodic lattice also has reflection symmetry, we have found that additional bound states can form in the positive energy continuum if their wave-functions are spatially antisymmetric.  When the reflection symmetry is broken, these states become long-lived quasibound states.  Although the analysis described here focused on a 1D periodic lattice (with finite width) in 2D space, we expect that similar BICs can also form in the energy continuum of 2D  periodic lattices, especially if they have additional symmetries.

\vspace{0.3cm}

\noindent {\bf ACKNOWLEDGEMENTS} The authors  thank the Robert A. Welch Foundation (Grant No. F-1051) for support of this work.

\section{Appendix A}

Compute matrix elements, taken with respect to the reaction region and asymptotic region eigenstates, of the coupling Hamiltonians.
\begin{eqnarray}
{\langle}{\phi}^{{\ell}}_{j} |{\hat Q}{\hat H}{\hat P}_B|{\psi}_{k^{{\ell}'},\nu}^{B} {\rangle}={\int_{-Na}^{Na}}dx_1{\int_{-L}^{L}}dz_1{\int_{-Na}^{Na}}dx_2{\int_{-\infty}^{-L}}dz_2~~~~~~~~~~~~~~~~~~~~~~~~~~~
~~~~~~~\nonumber\\
{\times}~{\langle}{\phi}^{\ell}_j|x_1,z_1{\rangle}~{\langle}x_1,z_1|{\delta}({\hat z}+L)~{\hat {\partial}}_{{z}}|x_2,z_2{\rangle}{\langle}x_2,z_2|{\psi}_{k^{\ell '}_{\nu},{\nu}}^B{\rangle}~~~~~~~~~~~
~~~~~\nonumber\\
=\frac{1}{2}{\delta}_{\ell,\ell '} \frac{1}{2Na}{\int_{-Na}^{Na}}dx_1{\bigg[}~{\bigg(}{\sum_{m={-\infty}}^{\infty}}{\sum_{n=0}^{\infty}}C^{j,{\ell}*}_{m,n}{\rm e}^{-iK_{\ell}x_1}{\rm e}^{-im{\pi}x_1/a}{\xi}_{n}(-L){\bigg)}~~~~~~~~~~~~~\nonumber\\
{\times}{\rm e}^{iK_{\ell}x_1}{\rm e}^{i{\nu}{\pi}x_1/a} {\bigg]}\frac{d{\xi}_{k_{\nu}}^B}{dz}{\bigg|}_{z=-L}
=\frac{1}{2}{\delta}_{\ell,\ell '}{\Phi}^{\ell *}_{j,\nu}(-L)~\frac{d{\xi}_{k_{\nu}}^B}{dz}{\bigg|}_{z=-L}~~~~~~~\end{eqnarray}

where
\begin{eqnarray}
{\Phi}^{\ell *}_{j,\nu}(-L)=\frac{1}{2Na}{\int_{-Na}^{Na}}dx_1~{\bigg(}{\sum_{m={-\infty}}^{\infty}}{\sum_{n=0}^{\infty}}C^{j,{\ell}*}_{m,n}{\rm e}^{-i(K_{\ell}+m{\pi}/a)x_1}{\xi}_{n}(-L){\bigg)}{\rm e}^{i(K_{\ell}+{\nu}{\pi}/a)x_1}
\end{eqnarray}

Similarly,

\begin{eqnarray}
{\langle}{\phi}^{{\ell}}_{j} |{\hat Q}{\hat H}{\hat P}_T|{\psi}_{k^{{\ell}'},\nu}^{T} {\rangle}={\int_{-Na}^{Na}}dx_1{\int_{-L}^{L}}dz_1{\int_{-Na}^{Na}}dx_2{\int_{L}^{\infty}}dz_2~~~~~~~~~~~~~~~~~~~~~~~\nonumber\\
{\times}~{\langle}{\phi}^{\ell}_j|x_1,z_1{\rangle}~{\langle}x_1,z_1|{\delta}({\hat z}-L)~{\hat {\partial}}_{{z}}|x_2,z_2{\rangle}{\langle}x_2,z_2|{\psi}_{k^{\ell '}_{\nu},{\nu}}^T{\rangle}~~~~~~~~~~
~~~~~~\nonumber\\
=\frac{1}{2}{\delta}_{\ell,\ell '} \frac{1}{2Na}{\int_{-Na}^{Na}}dx_1{\bigg[}~{\bigg(}{\sum_{m={-\infty}}^{\infty}}{\sum_{n=0}^{\infty}}C^{j,{\ell}*}_{m,n}{\rm e}^{-iK_{\ell}x_1}{\rm e}^{-im{\pi}x_1/a}{\xi}_{n}(L){\bigg)}~~~~~~\nonumber\\
{\times}{\rm e}^{iK_{\ell}x_1}{\rm e}^{i{\nu}{\pi}x_1/a} {\bigg]}\frac{d{\xi}_{k_{\nu}}^T}{dz}{\bigg|}_{z=L}~
=\frac{1}{2}{\delta}_{\ell,\ell '}{\Phi}^{\ell *}_{j,\nu}(L)~\frac{d{\xi}_{k_{\nu}}^T}{dz}{\bigg|}_{z=L}~~~~~~
\end{eqnarray}

where
\begin{eqnarray}
{\Phi}^{\ell *}_{j,\nu}(L)=\frac{1}{2Na}{\int_{-Na}^{Na}}dx_1~{\bigg[}{\bigg(}{\sum_{m={-\infty}}^{\infty}}{\sum_{n=0}^{\infty}}C^{j,{\ell}*}_{m,n}{\rm e}^{-i(K_{\ell}+m{\pi}/a)x_1}{\xi}_{n}(L){\bigg)}{\rm e}^{i(K_{\ell}+{\nu}{\pi}/a)x_1}{\bigg]}
\end{eqnarray}
\section{Appendix B}

We have constructed the reaction region eigenstates assuming zero-slope boundary conditions at $z={\pm}L$, where $L=3$.  One can show that, with this boundary condition, the W-E method gives an exact result for the scattering matrix of a 1D system.  For example, for a 1D system consisting of a potential well $V(z)=-V_0$ for $-3{\leq}z{\leq}3$ and $V(z)=0$ otherwise, the exact solution for the reaction matrices are $R_{BB}=\frac{Cot[2Lk_0]}{k_0}$ and $R_{BT}=\frac{Csc[2Lk_0]}{k_0}$, where $k_0=\sqrt{\frac{2m(E+V_0)}{{\hbar}^2}}$.   W-E theory gives
\begin{eqnarray}
R_{BB}(E)=\frac{2L}{{\pi}^2y^2} + ~\frac{4L}{{\pi}^2} {\sum_{n=1}^{\infty}}\frac{1}{(y^2-n^2)} ~~~~~~~~~~~~\nonumber\\
R_{BT}(E)=\frac{2L}{{\pi}^2y^2} + ~\frac{4L}{{\pi}^2} {\sum_{n=1}^{\infty}}\frac{(-1)^n}{(y^2-n^2)}.~~~~~~~~~~
\end{eqnarray}
where $y=\frac{2Lk_0}{\pi}$.  These are exact series expansions for $R_{BB}=\frac{Cot[2Lk_0]}{k_0}$ and $R_{BT}=\frac{Csc[2Lk_0]}{k_0}$.

\newpage

\subsection{List of Figures}

\begin{figure}[!hp]
\centering
 \scalebox{.8}{\includegraphics{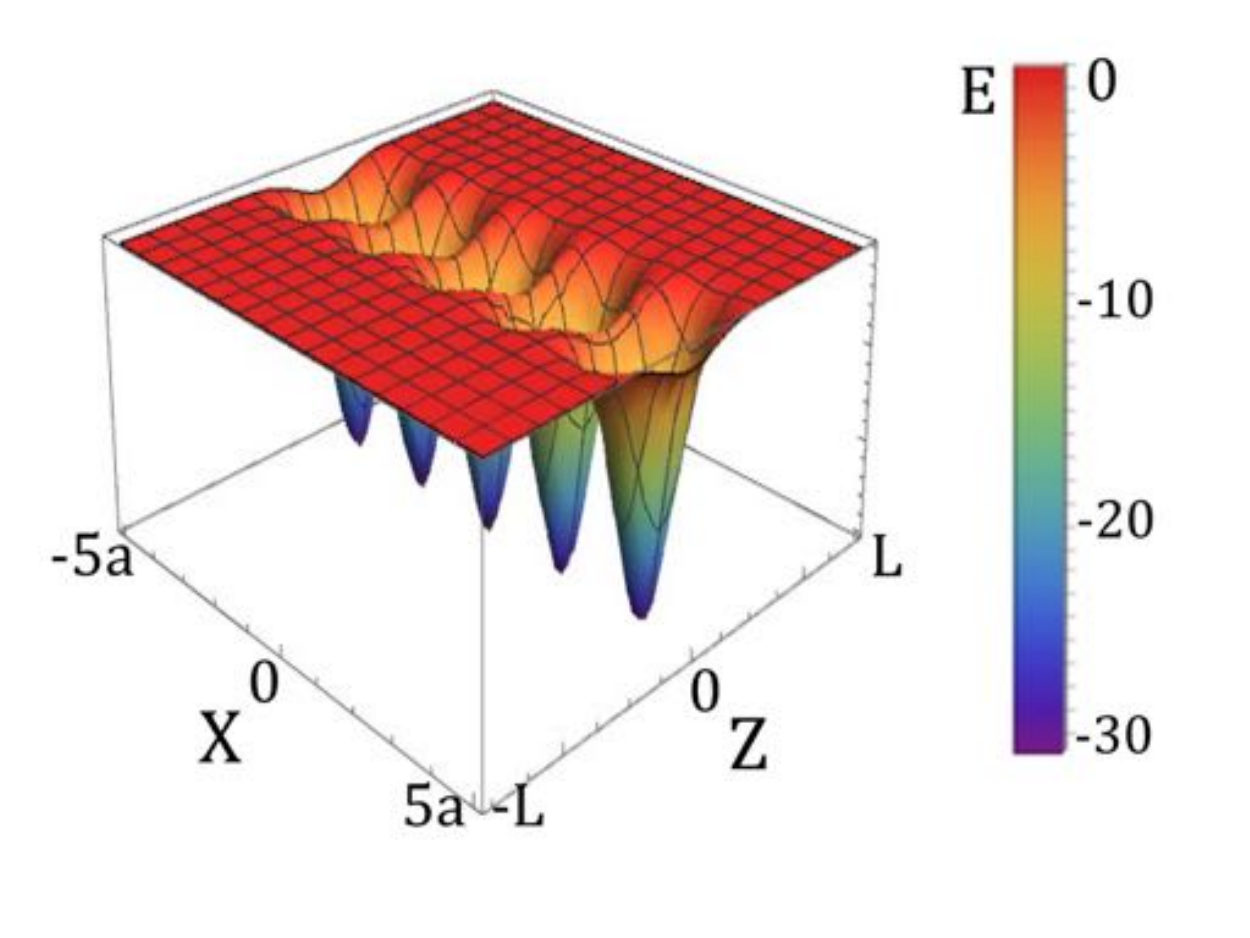}}
\caption{A finite segment of the potential energy in the reaction region for the case ${\beta}=0$. The $x$-direction shows five unit cells of the 1D infinite lattice, each with depth $V(0,0)=-29.84$. The $z$-direction shows the transverse width of the reaction region (here $L=3$). All quantities in dimensionless units. }
\label{fig:Lattice}
\end{figure}
%

\begin{figure}[!hp]
\centering
\scalebox{.8}{\includegraphics{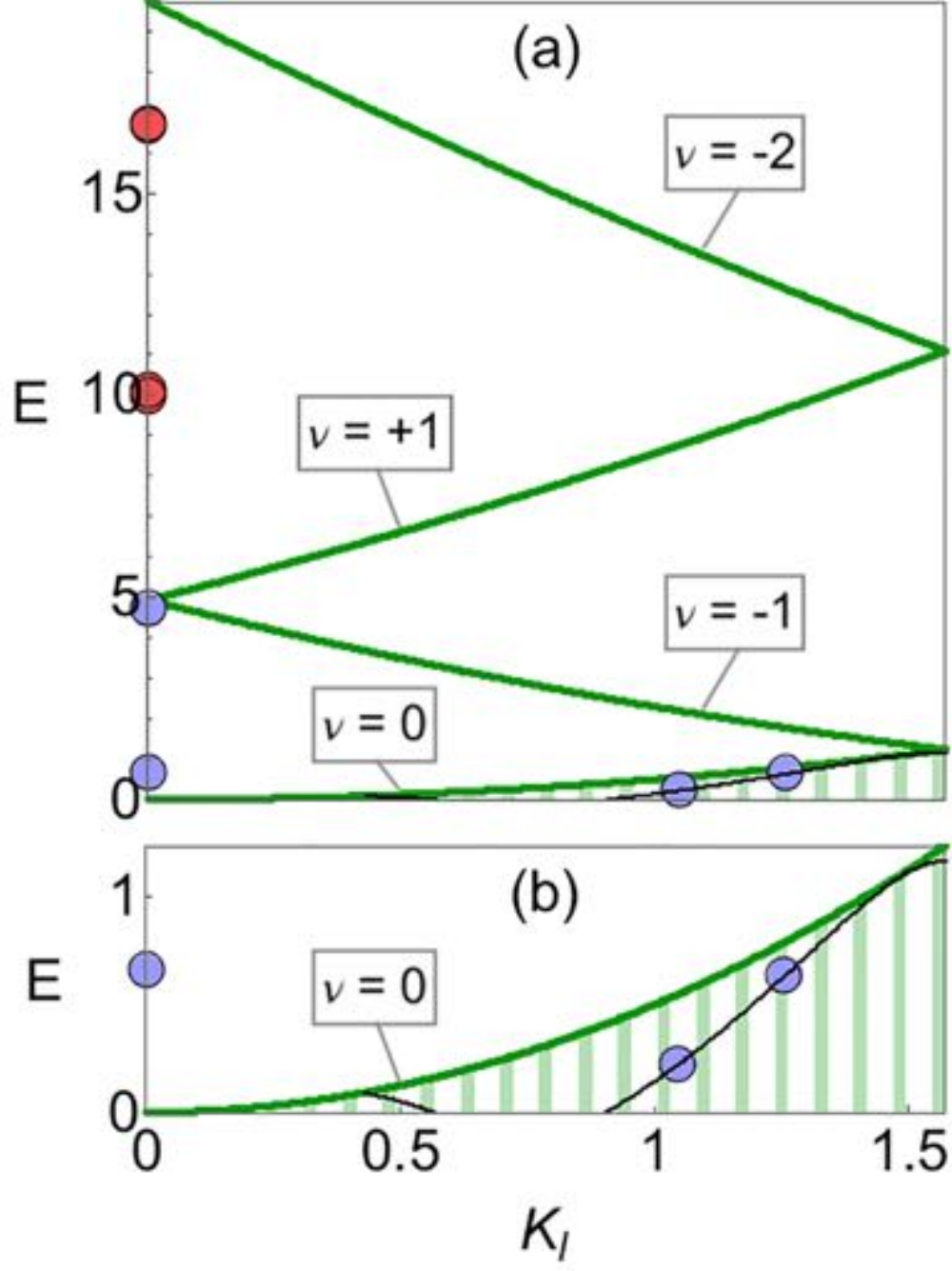}}
\caption{ Band structure of the asymptotic region when ${\beta}=0$  and  the $z$-component of the momentum $k_{\nu}^{\ell}=0$.  The index ${\nu}$ labels the scattering modes, ${\rm E}$ is the energy,   and ${\rm K}_{\ell}$ denotes the Bloch momentum in the interval $0{\leq}{\rm K}_{\ell}{\leq}\frac{\pi}{2}$. (a) $0{\leq}E{\leq}\frac{4{\pi}^2}{2}$: the lower two dots on the $\Gamma$ line ($K=0$) are BICs, the upper two dots on the $\Gamma$ line are long-lived quasibound states.  (b) $0{\leq}E{\leq}\frac{{\pi}^2}{8}$: the two dots (and all states) along the black line are BICs. All quantities in dimensionless units.}
\label{fig:bandasymp}
\end{figure}
%

\begin{figure}[!hp]
\centering
\scalebox{.8}{\includegraphics{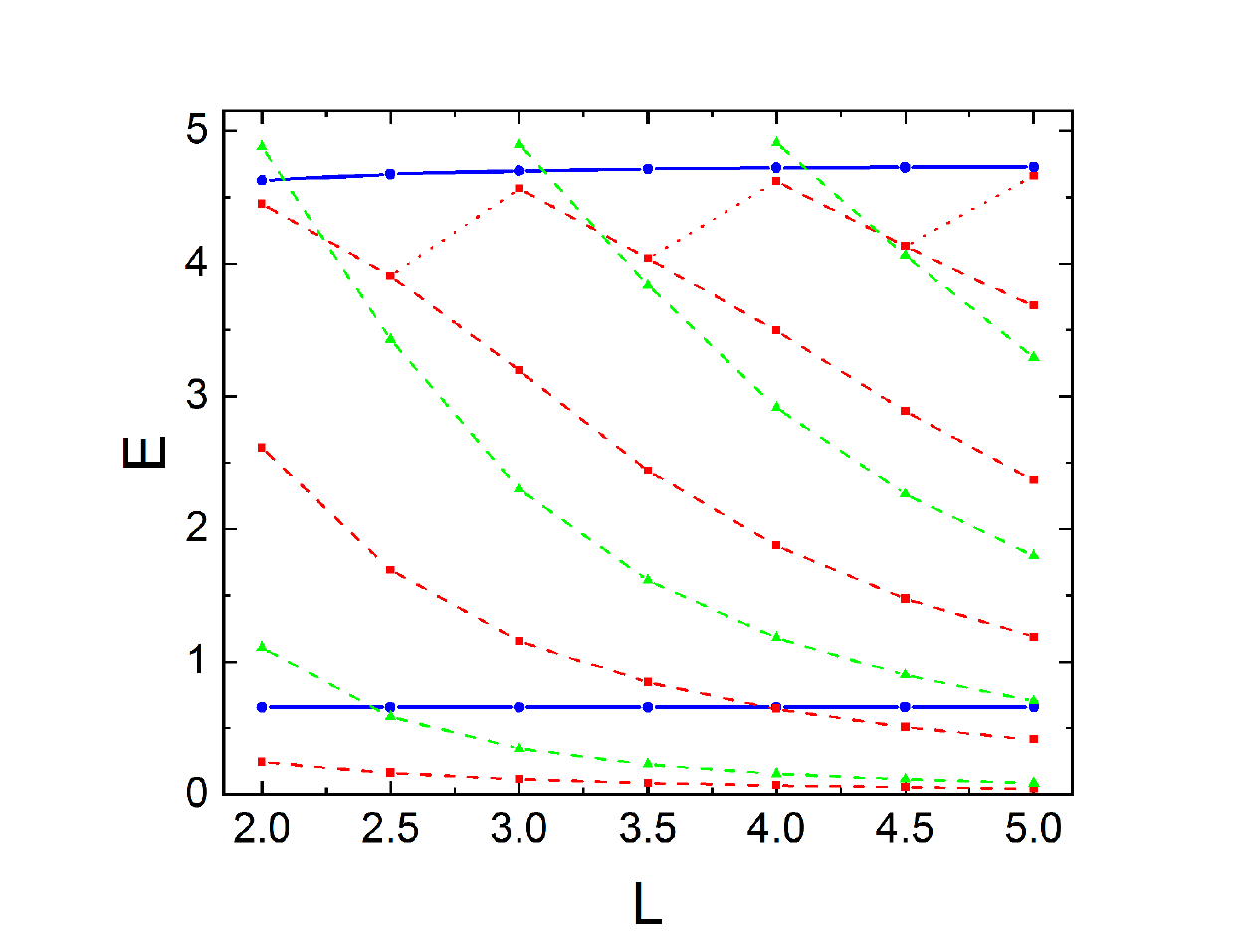}}
\caption{ The energy of  reaction region eigenstates in the first channel of the Gamma point ($0{\leq}E{\leq}\frac{{\pi}^2}{2}$) plotted as a function of $L$.  Two states (marked by the solid  line) have energy eigenvalues independent of the value of $L$ and are localized in the neighborhood of the potential (see Figs. \ref{fig:Ldependfigs}.a, \ref{fig:Ldependfigs}.b, \ref{fig:L5dependfigs}.a, and  \ref{fig:L5dependfigs}.b). As $L$ increases new states appear in the reaction region and have energy and structure that depends on $L$ (see Figs. \ref{fig:Ldependfigs}.c,  \ref{fig:Ldependfigs}.d, \ref{fig:L5dependfigs}.c, and  \ref{fig:L5dependfigs}.d). All quantities in dimensionless units. }
\label{fig:Ldepend}
\end{figure}
%

\begin{figure}[!hp]
\centering
\scalebox{.8}{\includegraphics{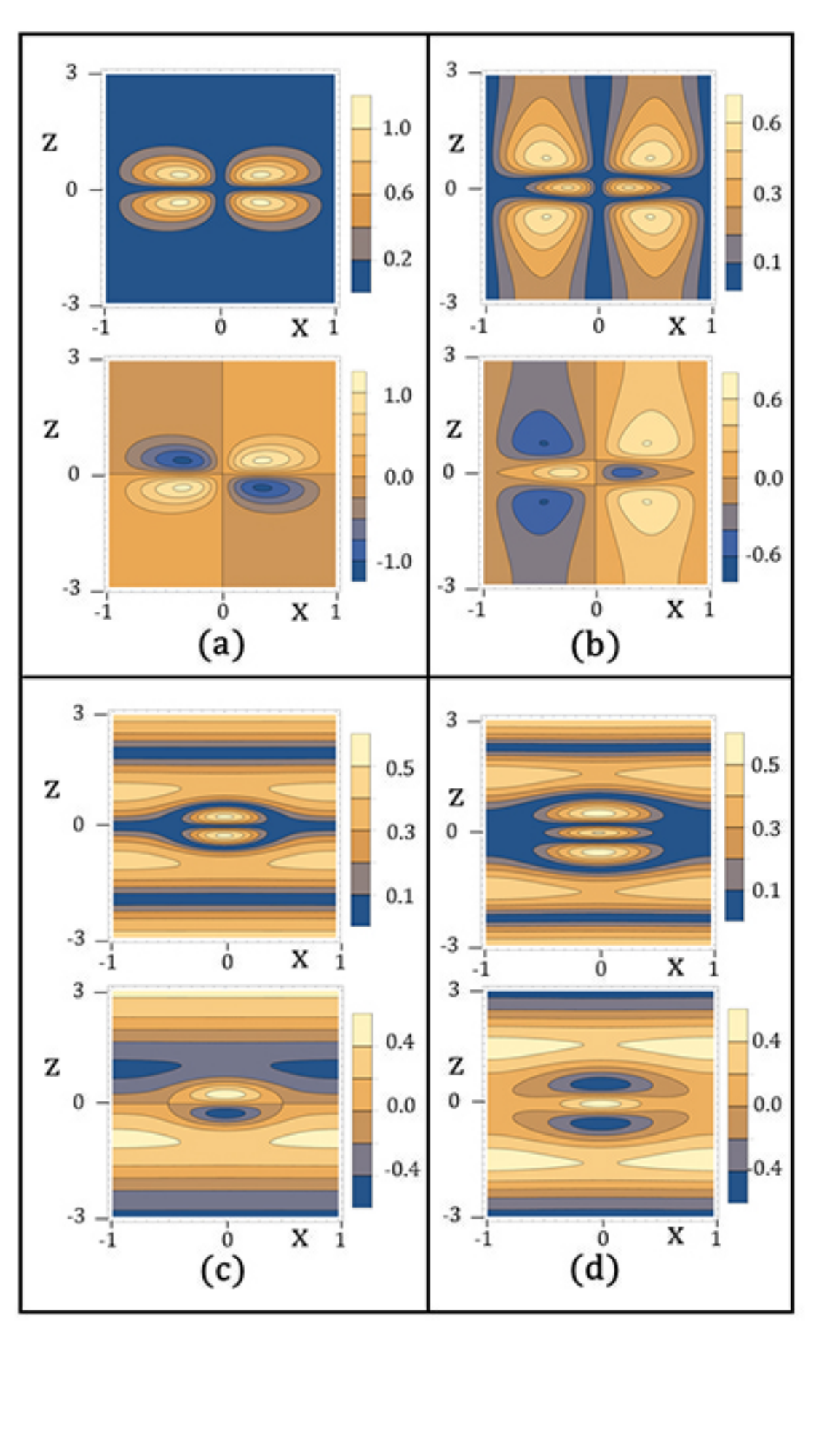}}
\caption{Reaction region eigenstates at the $\Gamma$ point for energies $0{\leq}E{\leq}\frac{{\pi}^2}{2}$,  $L=3$, and ${\beta}=0$. (a) L-independent (localized) state with energy $E=0.656436$ ( imaginary). (b) L-independent (localized) state with energy $E=4.69882$ (imaginary). (c) L-dependent state with energy $E=1.15902$ (real). 
 (d) L-dependent state with energy $E=2.30026$ (real).  The upper plot in each case is the absolute value of the state. The states in Figs. 4.a and 4.b are symmetry protected BICs. All quantities in dimensionless units.}
\label{fig:Ldependfigs}
\end{figure}
%

\begin{figure}[!hp]
\centering
\scalebox{.8}{\includegraphics{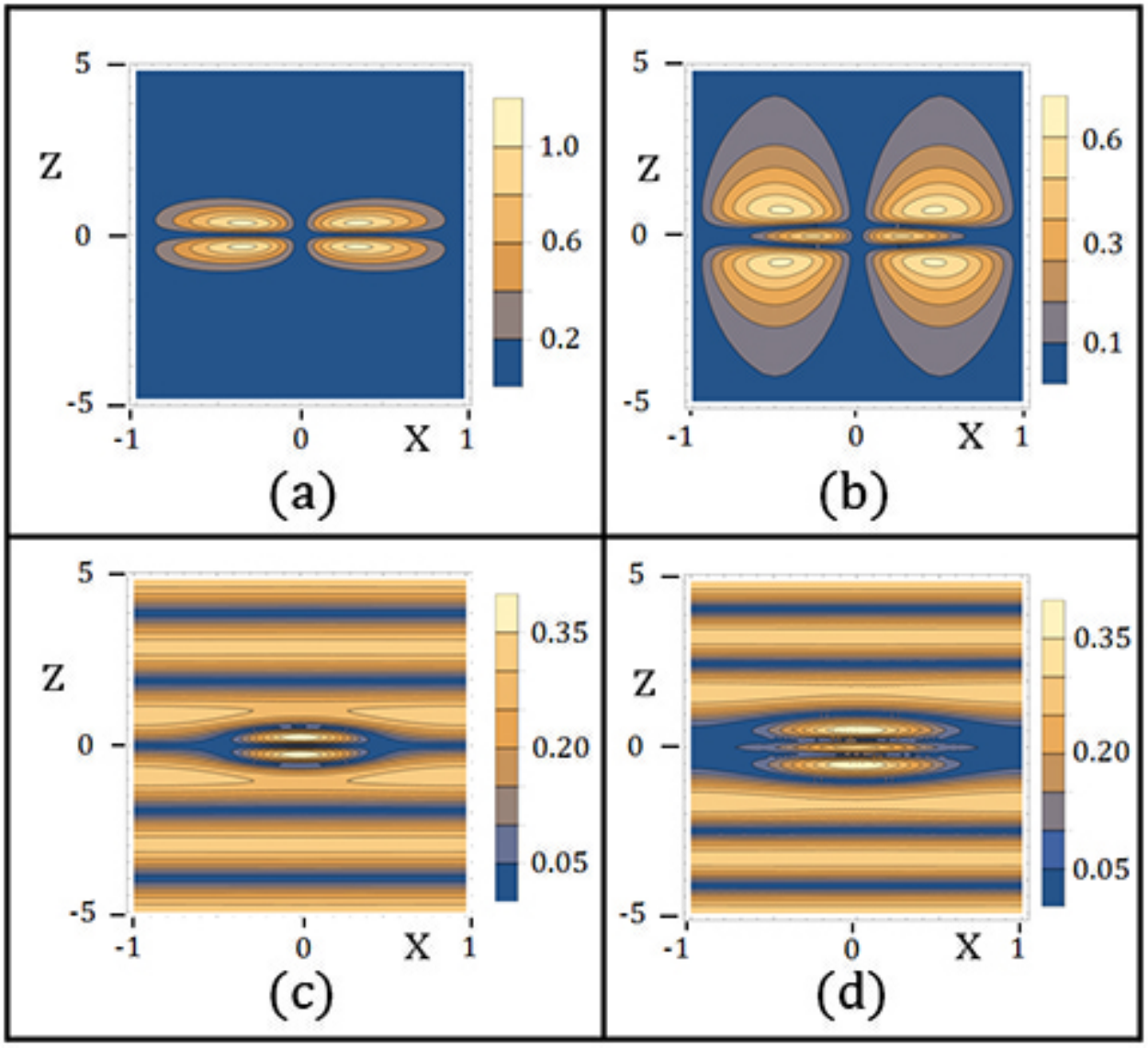}}
\caption{Reaction region eigenstates at the $\Gamma$ point for energies $0{\leq}E{\leq}\frac{{\pi}^2}{2}$. $L=5$, and ${\beta}=0$. (a) L-independent (localized) state with energy $E=0.65774$ (imaginary). (b) L-independent (localized) state with energy $E=4.7268$ (imaginary). (c) L-dependent state with energy $E=1.1907$ (real). 
 (d) L-dependent state with energy $E=2.797$ (real).  Only the absolute value is shown. The states in Figs. 5.a and 5.b are symmetry protected BICs. All quantities in dimensionless units.}\label{fig:L5dependfigs}
\end{figure}
%

\begin{figure}[!hp]
\centering
\scalebox{.8}{\includegraphics{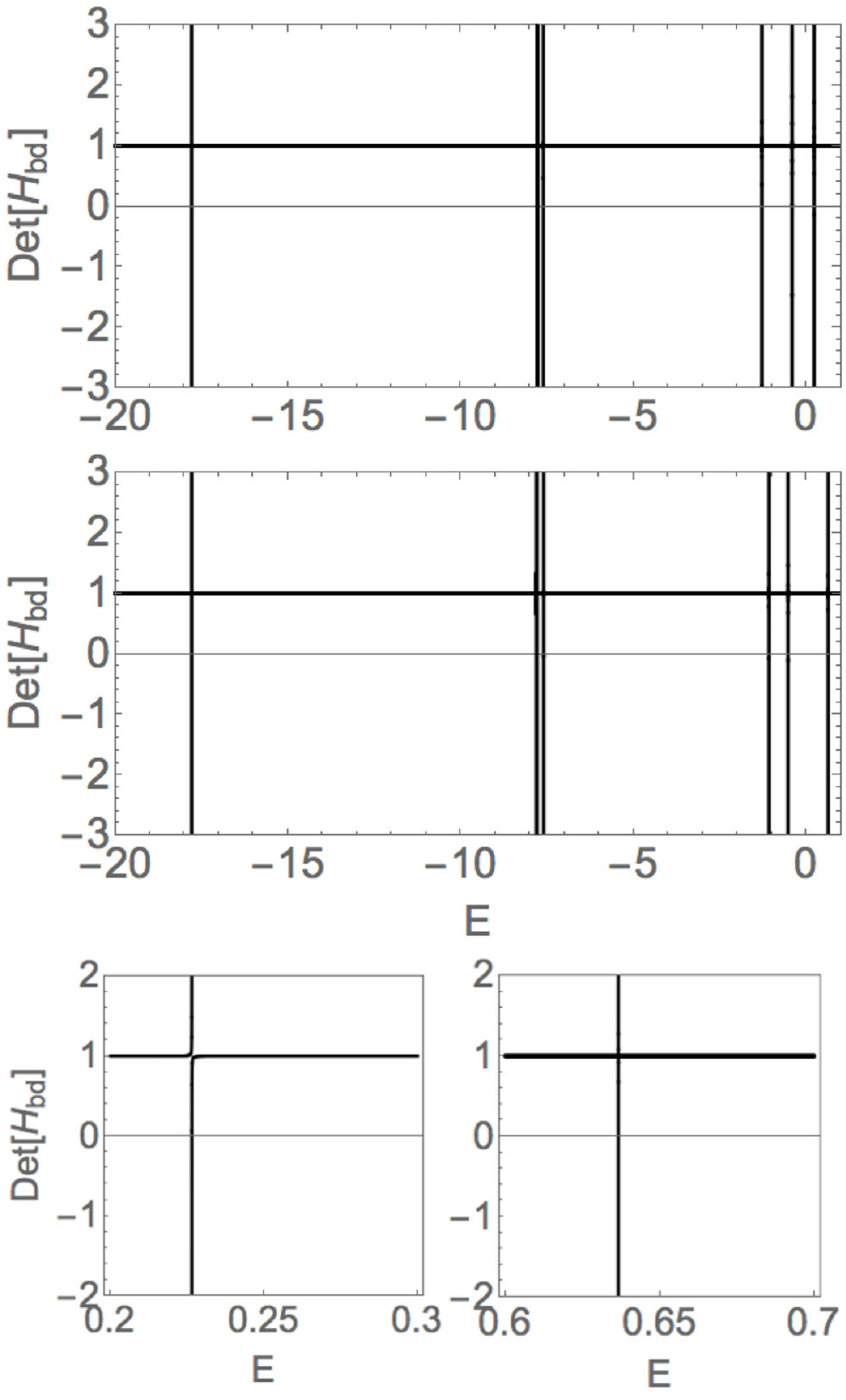}}
\caption{ Plots of ${\rm Det}[H_{bd}]$ as a function of energy $E$. ${\rm Det}[H_{bd}]$ is zero  at bound state energies below $E=K^2/2$. (a) Zeros of ${\rm Det}[H_{bd}]$ for the Bloch channel $K=\frac{{\pi}}{3}$.  (b) Zeros of ${\rm Det}[H_{bd}]$ for the Bloch channel $K=\frac{2{\pi}}{5}$.  (c) The zero of ${\rm Det}[H_{bd}]$ for  positive energy for Bloch channel $K=\frac{{\pi}}{3}$ indicates that a bound state exists in the energy continuum. (d) The zero of ${\rm Det}[H_{bd}]$ for positive energy  for Bloch channel $K=\frac{2{\pi}}{5}$ indicates that a bound state exists in the energy continuum. All quantities in dimensionless units.}
\label{fig:BlochBSspectrum}
\end{figure}
%

\begin{figure}[!hp]
\centering
\scalebox{.8}{\includegraphics{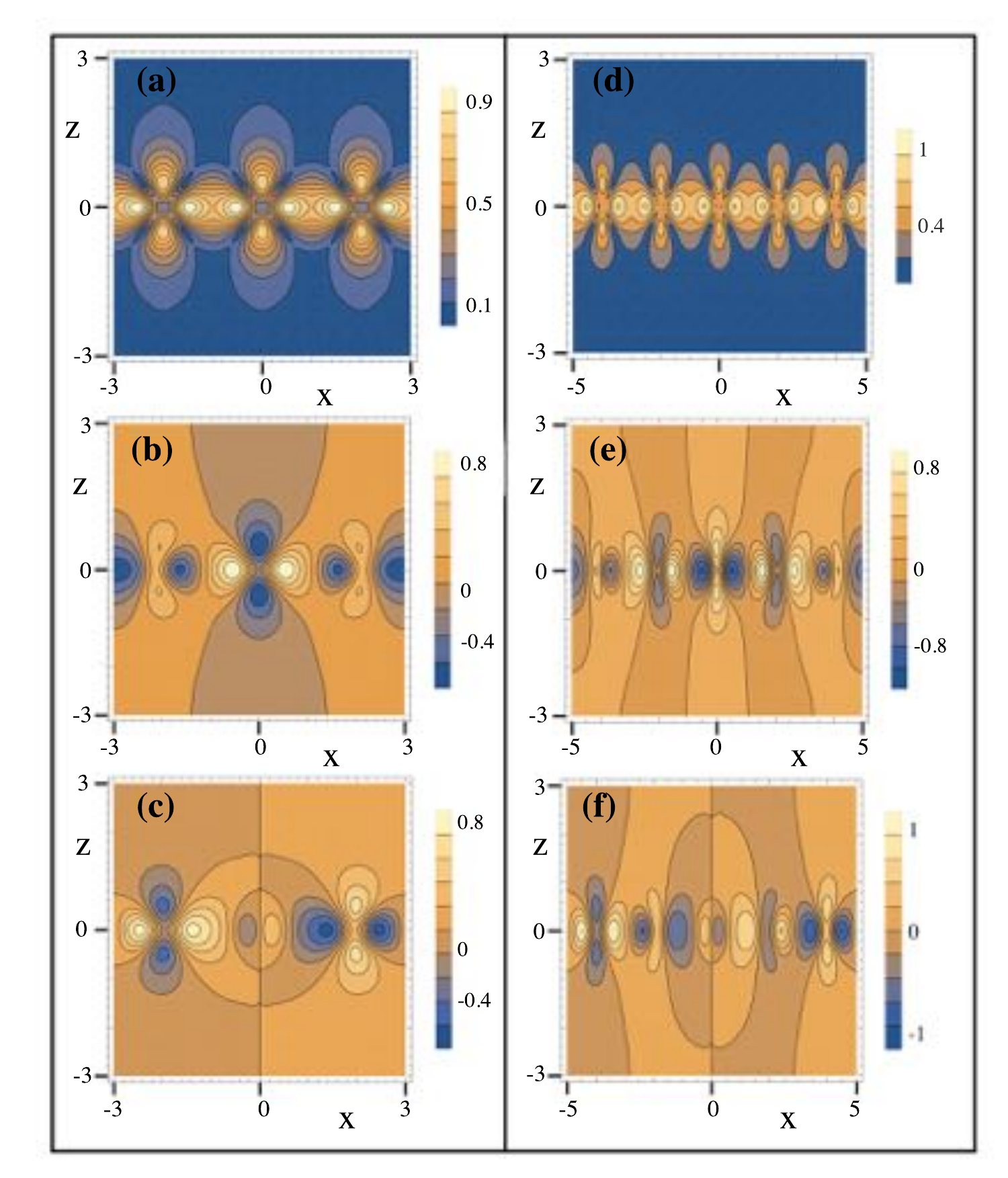}}
\caption{ Plots of reaction region Bloch states for $K=\frac{\pi}{3}$,~$E=0.2266$;  and for $K=\frac{2\pi}{5}$,~$E=0.6366$.  
 (a) $K=\frac{\pi}{3}$, Absolute value.  (b) $K=\frac{\pi}{3}$, Real part. (c) $K=\frac{\pi}{3}$, Imaginary part. (d) $K=\frac{2\pi}{5}$, Absolute value.  (e) $K=\frac{2\pi}{5} $, Real part. (f) $K=\frac{2\pi}{5}$ Imaginary part.  These states form BICs.   All quantities in dimensionless units.}
\label{fig:BlochBSplots}
\end{figure}
%

\begin{figure}[!hp]
\centering
\scalebox{.8}{\includegraphics{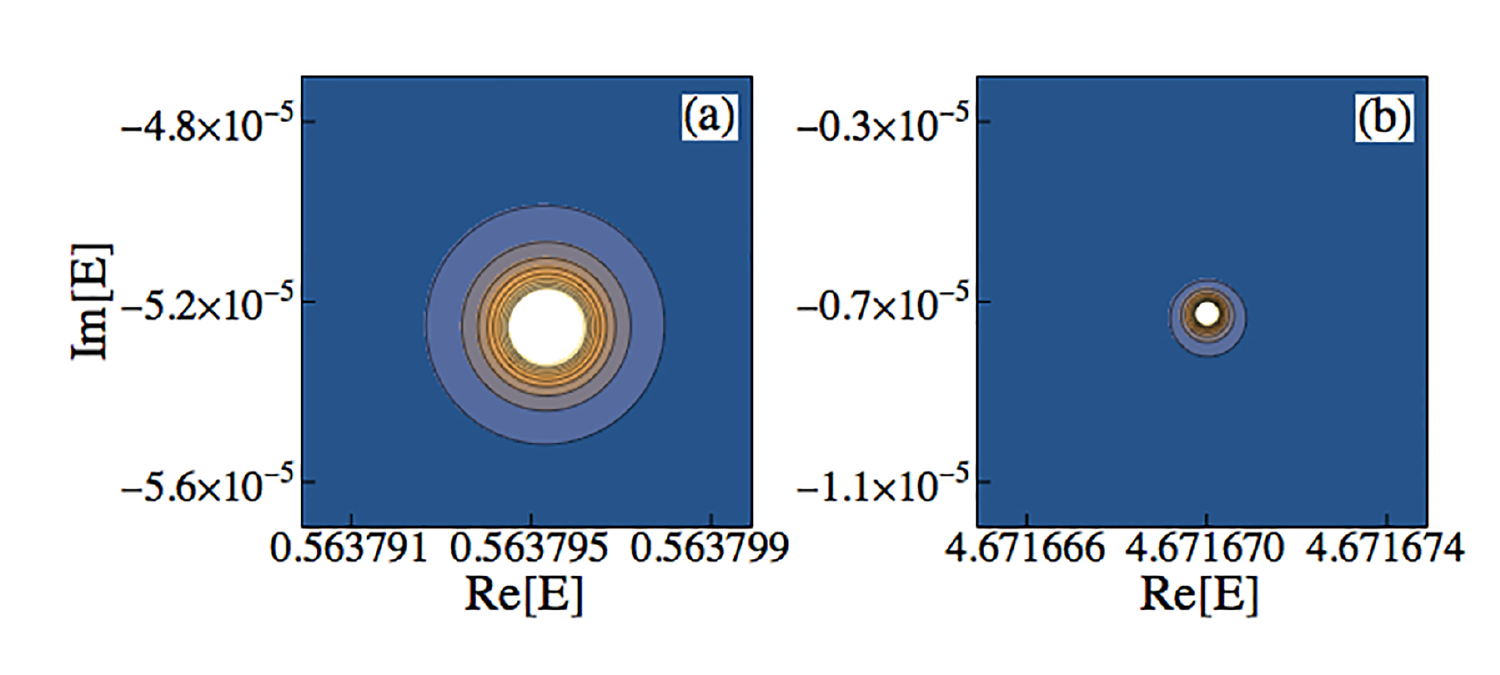}}
\caption{ Poles of the S-matrix at the $\Gamma$ point for ${\beta}=0.01$ and energy interval $0{\leq}E{\leq}\frac{{\pi}^2}{2}$.  All quantities in dimensionless units.}
\label{fig:BiCPoles}
\end{figure}
%

%
\begin{figure}[!hp]
\centering
\scalebox{.8}{\includegraphics{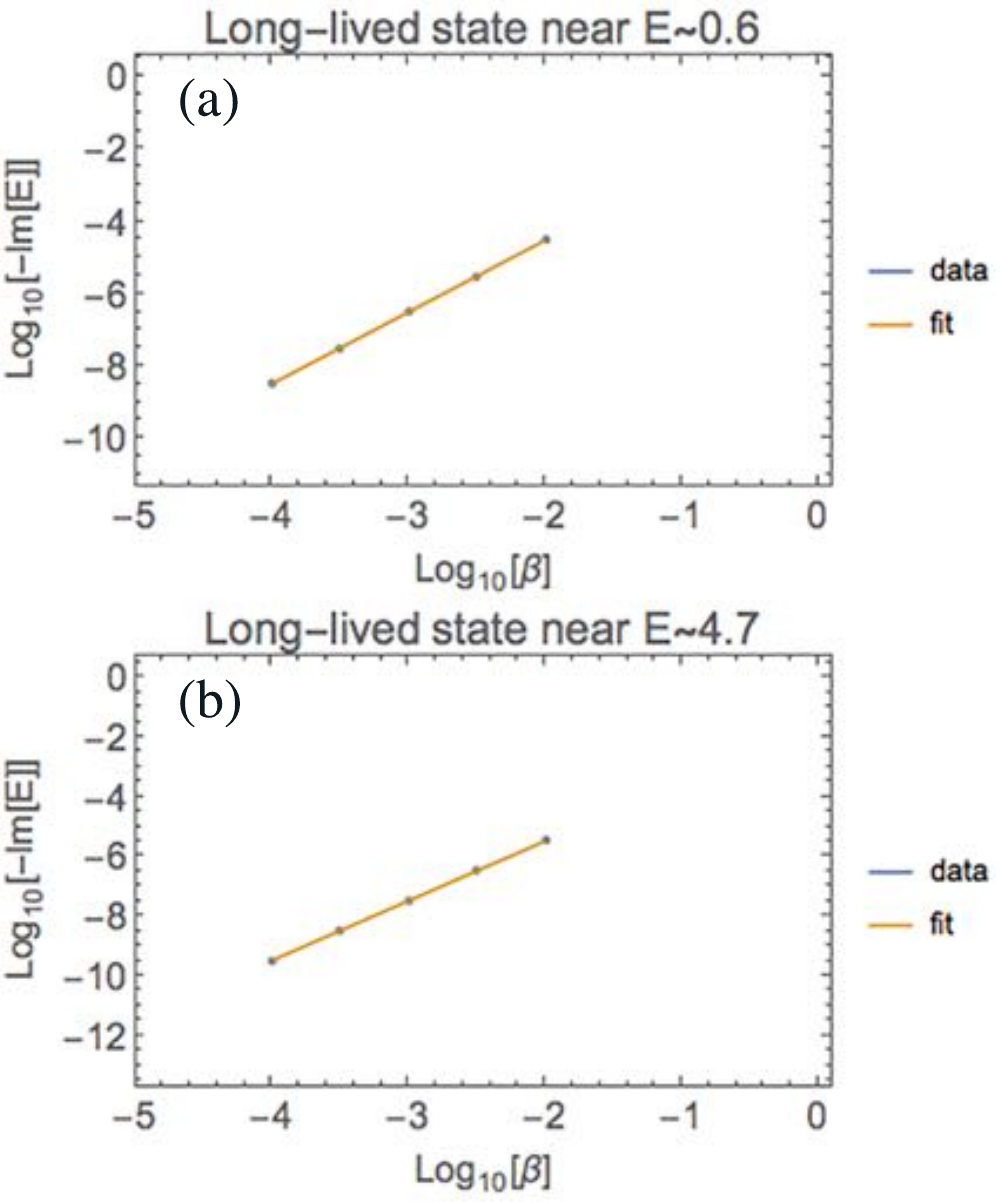}}
\caption{Lifetimes of the quasibound states that form for ${\beta}{\neq}0$ in the energy channel $0{\leq}E{\leq}\frac{{\pi}^2}{2}$ at the Gamma point. All quantities in dimensionless units.}
\label{fig:BiCLifetime}
\end{figure}
%

%
\begin{figure}[!hp]
\centering
\scalebox{.8}{\includegraphics{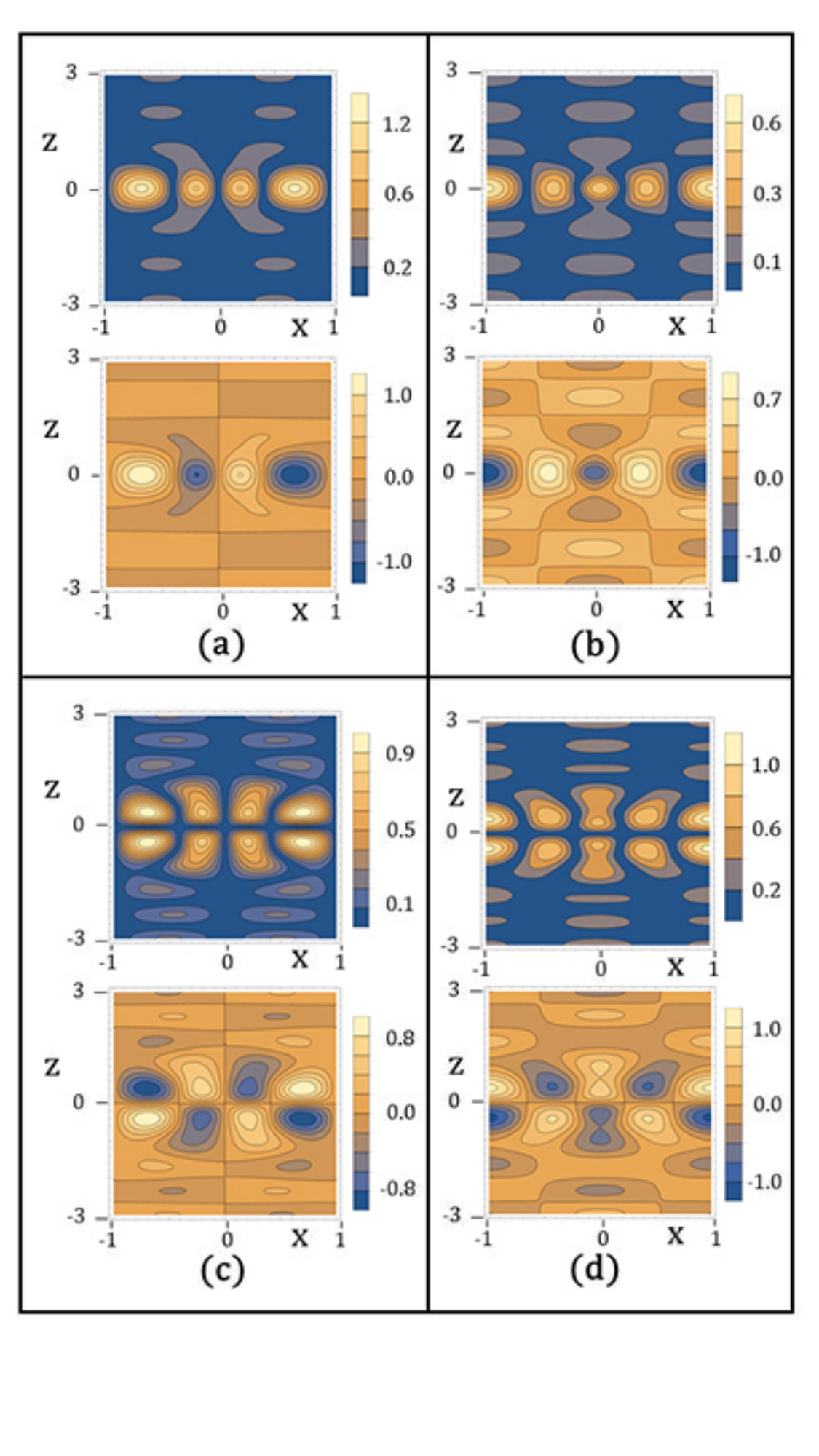}}
\caption{Plots of localized states for ${\beta}=0$  at the $\Gamma$  point for energies $\frac{{\pi}^2}{2}{\leq}E{\leq}\frac{4{\pi}^2}{2}$. These states are either pure real or pure imaginary. In each case the upper plot is the absolute value of the amplitude and the  lower plot is the amplitude.  (a) $E=9.9707$,  imaginary and antisymmetric. (b) $E=10.1369$, real and symmetric. (c)  $E=16.6538$, imaginary and antisymmetric. (d) $E=16.7215$, real and symmetric.    All quantities in dimensionless units.} 
\label{fig:UpGamStates}
\end{figure}
%

%
\begin{figure}[!hp]
\centering
\scalebox{.8}{\includegraphics{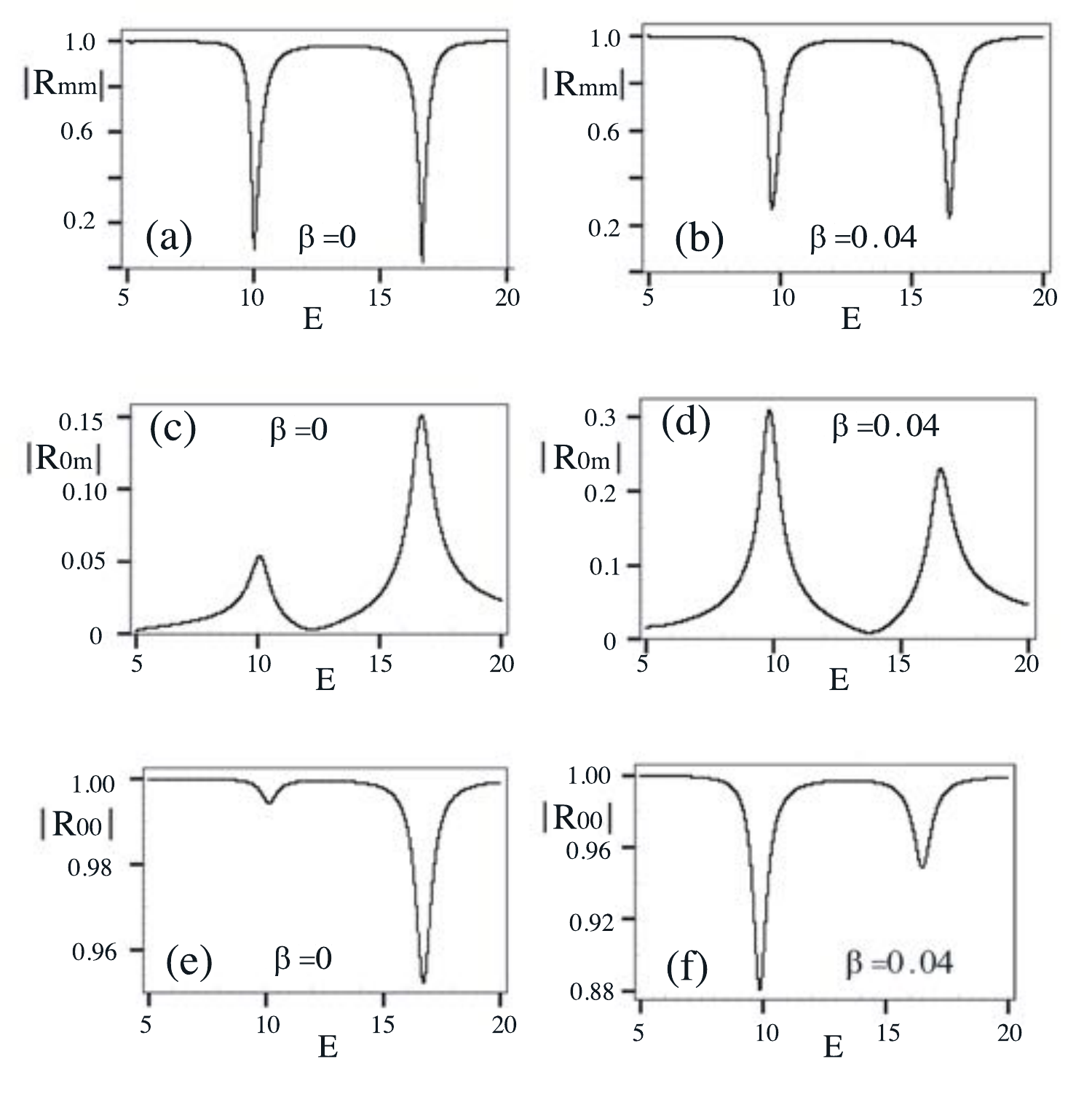}}
\caption{ Absolute value of reflection amplitudes for ${\beta}=0$ and for ${\beta}=0.04$ in the energy interval  $\frac{{\pi^2}}{2}{\leq}{\rm E}{\leq}\frac{4{\pi^2}}{2}$ at the $\Gamma$  point. (a) $|R_{mm}|$ for ${\beta}=0$.  (b) $|R_{mm}|$ for ${\beta}=0.04$. (c) $|R_{0m}|$ for ${\beta}=0$.  (d) $|R_{0m}|$ for ${\beta}=0.04$.  (e) $|R_{00}|$ for ${\beta}=0$. (f) $|R_{00}|$ for ${\beta}=0.04$.  (Only localized states were used to construct the reaction matrices for these plots.) All quantities in dimensionless units.}
\label{fig:UpGamReflec}
\end{figure}
%

%
\begin{figure}[!hp]
\centering
\scalebox{.8}{\includegraphics{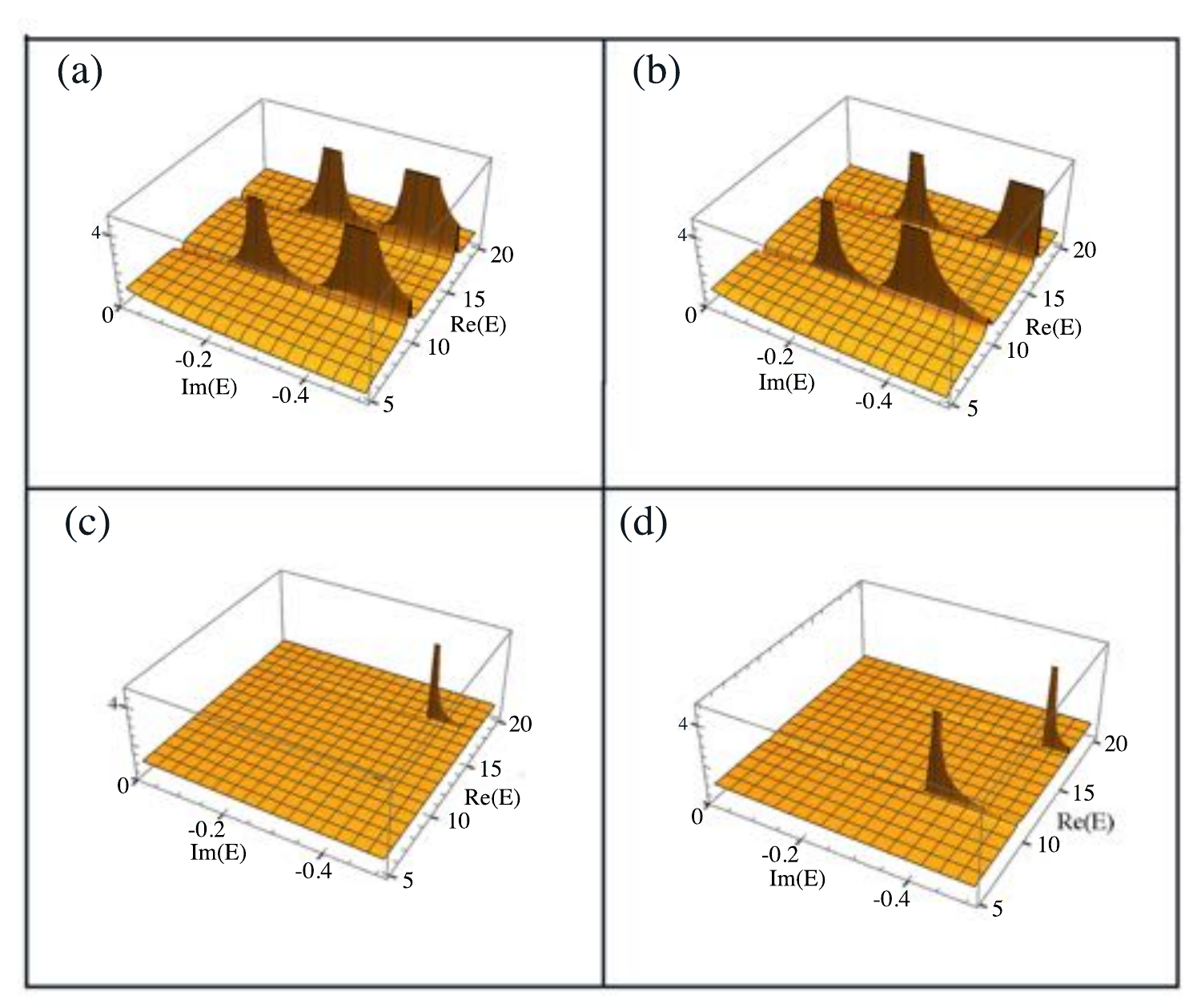}}
\caption{Poles  in the complex energy plane for reflection amplitudes in the energy interval  $\frac{{\pi^2}}{2}{\leq}{\rm Re(E)}{\leq}\frac{4{\pi^2}}{2}$ at the $\Gamma$  point. (a) Poles of  $R_{m,m}$ for ${\beta}=0$.  (b) Poles of  $R_{m,m}$ for ${\beta}=0.4$.  (c) Poles of reflection amplitude $R_{0,0}$ for ${\beta}=0$.  (d) Poles of reflection amplitude $R_{0,0}$ for ${\beta}=0.4$. (Only localized states were used to construct the reaction matrices for these plots.)  All quantities in dimensionless units.}
\label{fig:UpGamPoles}
\end{figure}

\end{document}